%% file: main.tex
\documentclass[reprint,
 amsmath,amssymb,
 aps,
prc,
raggedbottom,
]{revtex4-2}

\usepackage[normalem]{ulem}
\usepackage{graphicx}% Include figure files
\usepackage{dcolumn}% Align table columns on decimal point
\usepackage{bm}% bold math
\usepackage{epstopdf}
\usepackage{calrsfs}
\usepackage{calligra}
\usepackage{miama}
\usepackage{longtable}
\usepackage{multirow}
\usepackage{bm}
\usepackage{tabularx}
\usepackage{ltablex}
\usepackage{booktabs}

\usepackage[mathscr]{euscript}
\usepackage{breqn}

\usepackage{xcolor}

\usepackage{soul}

\usepackage{verbatim}
\usepackage{adjustbox}

\begin{document}
%\preprint{APS/123-QED}

\title{ Analysis of two-proton transfer in the $^{40}$Ca($^{18}$O,$^{20}$Ne)$^{38}$Ar reaction at 270 MeV incident energy}

\begin{abstract}

Two-nucleon transfer reactions are essential tools to investigate specific features of the nuclear structure such as the correlation among valence particles in the transfer process. Besides, transfer reactions may be an important channel to take into account in charge exchange processes since they can represent a competing contribution to the final cross section. 
The two-proton pickup transfer reaction $^{40}\mathrm{Ca}(^{18}\mathrm{O},^{20}\mathrm{Ne})^{38}\mathrm{Ar}$ has been measured at 270 MeV and the angular distributions for transitions to different excited states extracted. 
This work shows the analysis of the data performed by finite-range coupled reaction channel and coupled channel Born approximation methods. Extensive shell-model calculations are performed to derive the one- and two-proton spectroscopic amplitudes for the projectile and target overlaps. The role of the simultaneous and sequential two-proton transfer mechanisms to populate the measured final states or groups of states, mainly characterized by a high collectivity, is also discussed.      

\end{abstract}

\author{
J. L. Ferreira$^1$, 
D. Carbone$^2$, 
M. Cavallaro$^{2*}$, 
N. N. Deshmukh$^{2,3}$, 
C. Agodi$^2$, 
G. A. Brischetto$^{2,4}$, 
S. Calabrese$^2$, 
F. Cappuzzello$^{2,4}$, 
E. N. Cardozo$^1$, 
I. Ciraldo$^{2,4}$, 
M. Cutuli$^{2,4}$, 
M. Fisichella$^2$, 
A. Foti$^{4,5}$, 
L. La Fauci$^{2,4}$, 
O. Sgouros$^2$, 
V. Soukeras$^2$,  
A. Spatafora$^{2,4}$, 
D. Torresi$^2$, 
J. Lubian$^1$ 
for the NUMEN collaboration
}

\affiliation{$^1$Instituto de F\'{i}sica$,$ Universidade  Federal Fluminense$,$ Niter\'{o}i$,$ Rio de Janeiro$,$ 24210-340 Brazil}
\affiliation{$^2$Istituto Nazionale di Fisica Nucleare - Laboratori Nazionali del Sud$,$ via S. Sofia 62$,$ 95123$,$ Catania$,$ Italy}
\affiliation{$^3$School of Sciences$,$ Auro University$,$ Surat$,$ India}
\affiliation{$^4$Dipartimento di Fisica e Astronomia "Ettore Majorana"$,$ University of Catania$,$ via Santa Sofia 64$,$ 95123$,$ Catania Italy}

\maketitle 

\section{\label{intro} INTRODUCTION}

In the last years, the study of the nuclear matrix element (NME) involved in the description of neutrinoless double beta decay (0$\nu\beta\beta$) rate has being intensified mainly because their accurate knowledge is a crucial piece of information in the determination of the neutrino absolute mass scale, providing that 0$\nu\beta\beta$ decay rate is given \cite{Ejiri19, Engel17}.\\

The NUMEN (NUclear Matrix Elements for Neutrinoless double beta decay)  project \cite{CAC18} proposes to measure the cross sections of heavy-ion induced double charge exchange (DCE) reactions and of the competing transfer channels involving same projectile and target.
The idea is that DCE and 0$\nu\beta\beta$ have many common aspects, especially because the initial and final nuclear states involved in the transitions are the same \cite{Len19}. Thus, information on NME extracted from DCE experimental measurements can give valuable constraints on the determination of 0$\nu\beta\beta$ NME.\\

Within the NUMEN project, the study of multi-nucleon transfer processes in similar dynamical conditions as the explored DCE reactions is an important tool to get selective information about the involved nuclear wave functions, including the mean-field dynamics and the correlations among nucleons. Such set of information is of interest for the complete description of the DCE reaction mechanism and the role of the competition with the direct meson exchange mechanism in DCE \cite{Len19, Bel20, San18}.\ 

Heavy-ion induced multi-nucleon transfer studies were boosted in recent years with the main aim to understand the role of one-step (simultaneous) and two-step (sequential) mechanism in the transfer process. Therefore, one and two-neutron transfer reactions were extensively investigated, as reported in Refs. \cite{CCB13,CBB14,PSR15,NatComm,ECL16,CLL17,CFC17,PSV17,CLL18, AGC18,LEL18,BCC18}. In those works, the comparison with theoretical results shows that the transfer mechanism depends also on the nuclear structure of the involved nuclei. States with low collectivity are preferably accessed by the one-step mechanism. On the other hand, states with high collectivity are accessed by a two-step mechanism. Therefore, the two-neutron transfer process depends on the degree of collectivity of the final nuclear states, that can break (or not) the correlation of the two transferred neutrons \cite{PSV17, satchler}.\\

The one- and two-proton transfer reaction mechanisms were also investigated in literature \cite{RSS95,PSR98,TES78,JRC02,BDK07,KMJ12,DMD80,CFC2020}. In the past, the agreement between the theoretical calculations and the experimental data was poor in the two-proton transfer reaction. Large scaling factors were needed to warrant good agreement between theory and data~\cite{EFH76,TES78,RGG91}. 
In Ref. \cite{JRC02}, the experimental data for the two-proton stripping reaction $^{90}$Zr($^{16}$O,$^{14}$C)$^{92}$Mo were described using an extreme cluster model (that assumes spectroscopic amplitudes equal to 1 for all the considered overlaps), where the two transferred nucleons are treated as a cluster, with only the
component with the two neutrons coupled to a zero intrinsic angular momentum participating in the transfer. However, the theoretical results were strongly dependent on the bound-state well radius used to derive the wave functions of the two protons.\\ 

The two-proton and two-neutron transfer cross sections to low-lying states populated in the $^{20}$Ne + $^{116}$Cd collision have been recently studied in \cite{CFC2020} stressing the importance of a complete quantum-mechanical treatment of the reaction dynamics and nuclear structure description.\\

An interesting reaction is the ($^{18}$O,$^{20}$Ne) two-proton pickup, which is analogous to the (n,$^3$He) reaction \cite{SFG72,RHK76,iaea}, with the advantage of the availability of high-quality stable beams. 
Furthermore, the use of oxygen projectiles in two-proton pickup reactions is in principle quite convenient for spectroscopic studies, thanks to a controlled description of the projectile  closed or almost closed-shell structure. 
However, the ($^{18}$O,$^{20}$Ne) reaction has been rarely employed, and mainly for nuclear mass measurements ~\cite{RHK76,ZGL97}. 
The lack of microscopic analysis tools for heavy-ion induced reactions has probably hindered its use for spectroscopic purposes \cite{SFG72} in the past.
In Ref. \cite{SFG72} the study of the $^{40}$Ca($^{18}$O,$^{20}$Ne)$^{38}$Ar two-proton pickup reaction at 48 MeV was reported. The shapes of the angular distributions were typical of a grazing reaction at energies near the Coulomb barrier. In that paper, the cross section angular distributions for the transitions to the ground and first 2$^+$ excited state of $^{38}$Ar and $^{20}$Ne were analyzed. A simplified cluster model was used for the di-proton transfer within distorted wave Born approximation (DWBA). 
The need to use a normalization factor in the calculated cross sections and the not proper description of the angular distribution shape, especially for the transition to the $2^+$ excited state of $^{20}$Ne, make the authors conclude that the sequential transfer mechanism should be included for a good description of the data.\\

Nowadays the impressive progresses in nuclear reaction theory open unprecedented opportunities to adopt such heavy-ion reactions as tools for the investigation of the nuclear structure and reaction mechanisms.\\

In the present work, we investigate the two-proton pickup reaction $^{40}$Ca($^{18}$O,$^{20}$Ne)$^{38}$Ar at 270 MeV incident energy for the first time.
This reaction is the first step of the multi-nucleon transfer routes leading to the same final states of the 
 $^{40}$Ca($^{18}$O,$^{18}$Ne)$^{40}$Ar reaction that has been studied as a pilot experiment \cite{CCA15} of the NUMEN research program.
Energy spectra and angular distributions of the transitions to the ground and excited states of the populated states have been measured and analysed for the first time.
 Coupled reaction channels (CRC) and coupled channels Born approximation (CCBA) formalism are used to describe the direct and sequential transfer mechanisms, respectively.\\

This paper is organized as follows: the experimental setup and data reduction procedure are reported in Sec. II; a brief description of the theoretical formalism used to perform two-proton transfer calculations is given in Sec. III; the theoretical analysis is described in Sec. III; the results are discussed in Sec. IV and the conclusions are given in Sec. V.

\section{\label{exp} EXPERIMENTAL SETUP AND DATA REDUCTION}

The $^{40}$Ca($^{18}$O,$^{20}$Ne)$^{38}$Ar two-proton transfer reaction was measured at the INFN-LNS laboratory in Catania. 
A beam of $^{18}\mathrm{O}^{4+}$ ions, extracted by the K800 Superconducting Cyclotron accelerator, bombarded a 280$\pm$14 $\mu$g/cm$^2$ Ca target at 270 MeV incident energy. 
The Ca material for the target was evaporated on a carbon backing 25 $\mu$g/cm$^2$ thick, and then covered by a layer of 15 $\mu$g/cm$^2$ carbon and maintained in vacuum to reduce oxidation processes.
The ejectiles produced in the collisions were momentum-analysed by the MAGNEX large acceptance spectrometer \cite{CAC16} and detected by its focal plane detector (FPD) \cite{Cav12, Tor20}. The optical axis of the spectrometer was located at $\theta_{\mathrm{opt}} = +4^\circ$. 
The MAGNEX quadrupole and dipole magnetic fields were set in order that the $^{18}$O$^{8+}$ stripped beam, after passing through the target and the magnets, reaches a region beside the FPD but external to it. Since the magnetic rigidity ($B\rho$) of the $^{18}$O$^{8+}$ beam is higher than the one of the ejectiles of interest ($^{20}$Ne$^{10+}$), a specifically designed Faraday cup was placed in the high-$B\rho$ region aside the FPD, to stop the beam and measure the incident charge in each run. 
Thanks to the large angular acceptance of MAGNEX, an angular range of $0^\circ < \theta_{lab} < +8^\circ$ in the laboratory frame was explored in a unique angular setting, corresponding to scattering angles in the center of mass $0^\circ < \theta_{c.m.} < 12^\circ$.\\  

\begin{figure}[htb!]
\centering
\includegraphics[scale=0.45]{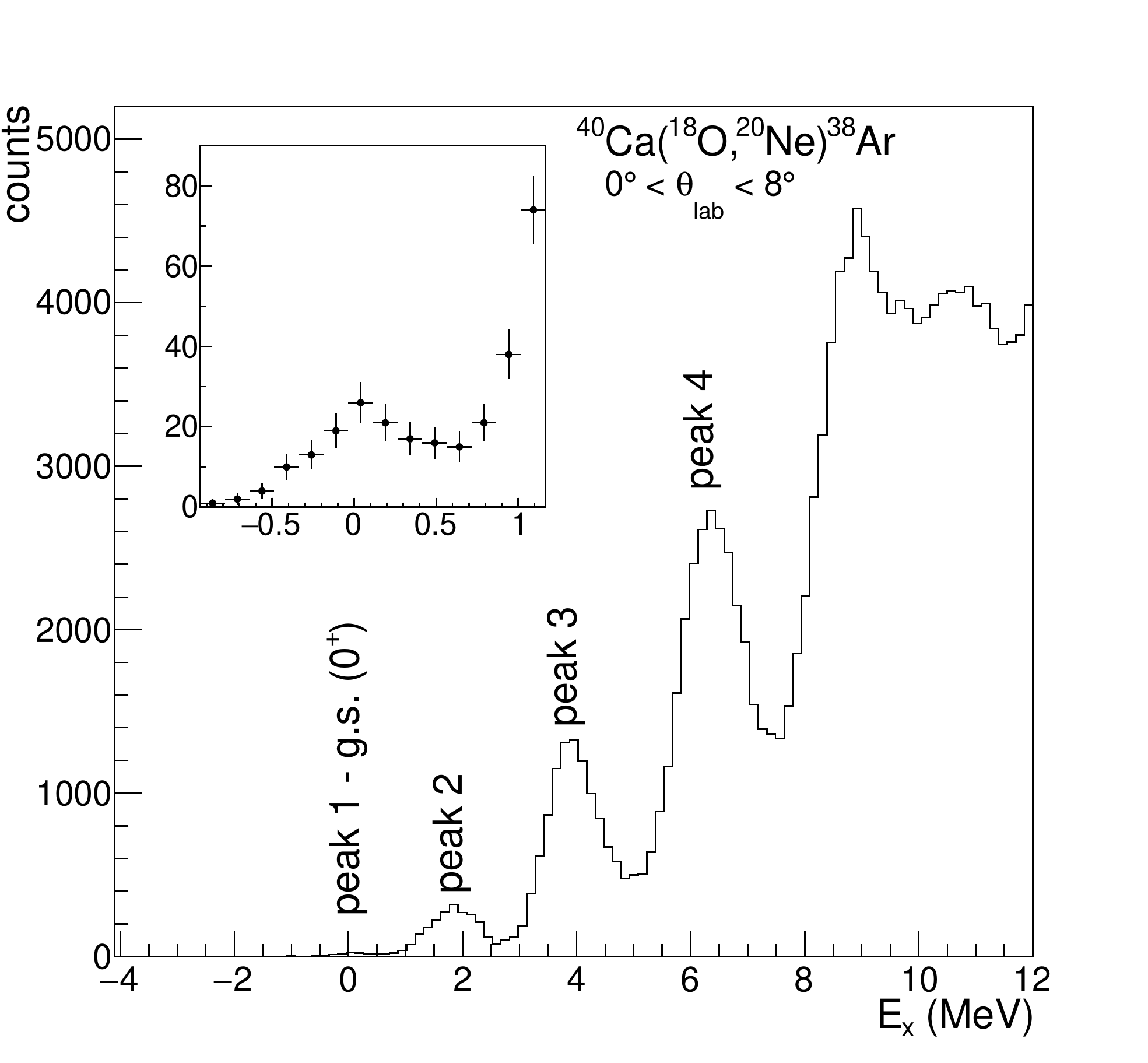} 
\caption {Excitation energy spectrum for the $^{40}$Ca($^{18}$O,$^{20}$Ne)$^{38}$Ar reaction at 270 MeV in the angular range $0^\circ < \theta_{lab} < +8^\circ$. In the inset a zoomed view of the ground state region is shown.}
\label{exp-fig1}
\end{figure}

The identification of the $^{20}$Ne ejectiles was performed using the technique described in Refs. \cite{Cav20, Cap10, Cal18, Cal20}. The positions and angles of the selected ions measured at the focal plane were used as input for a 10$^{th}$ order ray-reconstruction of the scattering angle $ \theta_{\mathrm{lab}}$ and excitation energy E$_x$ = Q$_0$ – Q (where Q$_0$ is the ground-to-ground state reaction Q-value) \cite{Cap11}.
The ray-reconstruction procedure also allows an accurate determination of the overall detection efficiency, fundamental to extract the absolute cross section from the collected event yields, as presented in Ref. \cite{Cav11}. A systematic error of about 10$\%$ in the cross section determination was estimated from the uncertainty in the target thickness and beam collection. It is not shown in the figures of the angular distributions as it is common to all the data points. \\ 

Fig.~\ref{exp-fig1} shows an example of the measured energy spectrum. The optimum Q-value for this reaction is around E$_x$ = 33 MeV, which explains the poor matching and thus the low cross section in the region at low excitation energy. On the top of this trend, different structures are visible in the spectrum. 
The first peak is the transition to the ground state of $^{20}$Ne and $^{38}$Ar, which is well isolated even if weakly populated. The other peaks are the result of groups of levels associated to both ejectile and residual nucleus excitation, not experimentally resolved due to the finite resolution (500 keV full width at half maximum). 
In particular, the second visible peak includes the transition to the $^{20}$Ne$_{\mathrm{g.s.}}(0^+)$ + $^{38}$Ar$_{2.17}(2^+)$ and $^{20}$Ne$_{1.63}(2^+)$ + $^{38}$Ar$_{\mathrm{g.s.}}(0^+)$ final states. The third and fourth peaks are a convolution of different final channels. See Section IV for a detailed description of the involved transitions.\\

\begin{figure} [h!]
\centering
\includegraphics[scale=0.42]{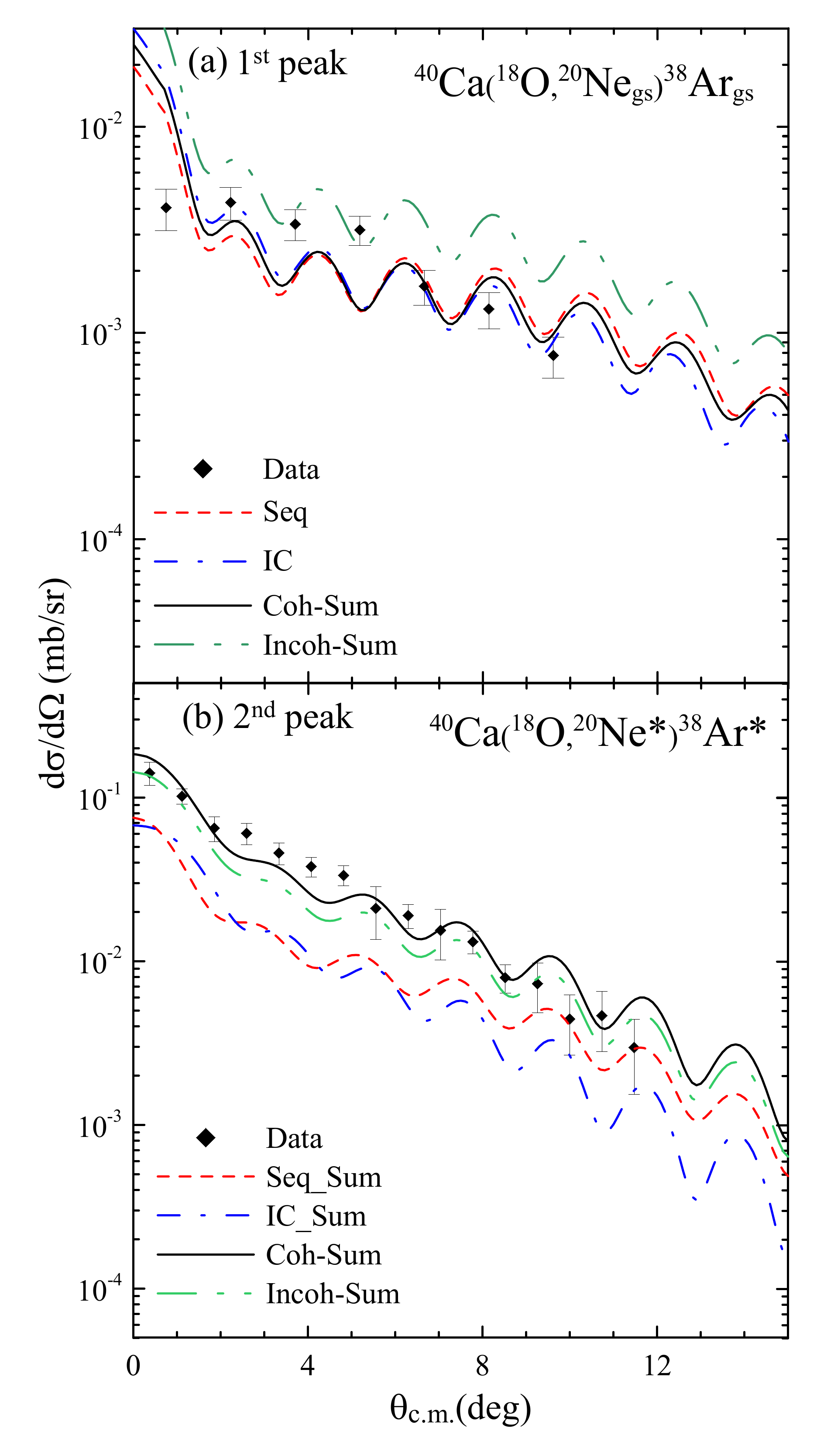} 
\caption {(color online) Comparison between the theoretical and experimental two-proton transfer angular distribution corresponding to: (a) the $^{20}\mathrm{Ne}_{\mathrm{gs}}(0^{+})+^{38}\mathrm{Ar}_{\mathrm{gs}}(0^{+})$ channel. (b) unresolved excited states concerning the second peak in Fig.~\ref{exp-fig1}. In both figures the contribution due to the simultaneous (IC) and sequential (Seq) transfer and the coherent (Coh) and incoherent (Incoh) sum of the two mechanisms are shown (see text). In (b) the sum of the different channels contributing to the cross section is plotted.}
\label{tr-fig2}
\end{figure}

\begin{figure} [h!]
\centering
\includegraphics[scale=0.42]{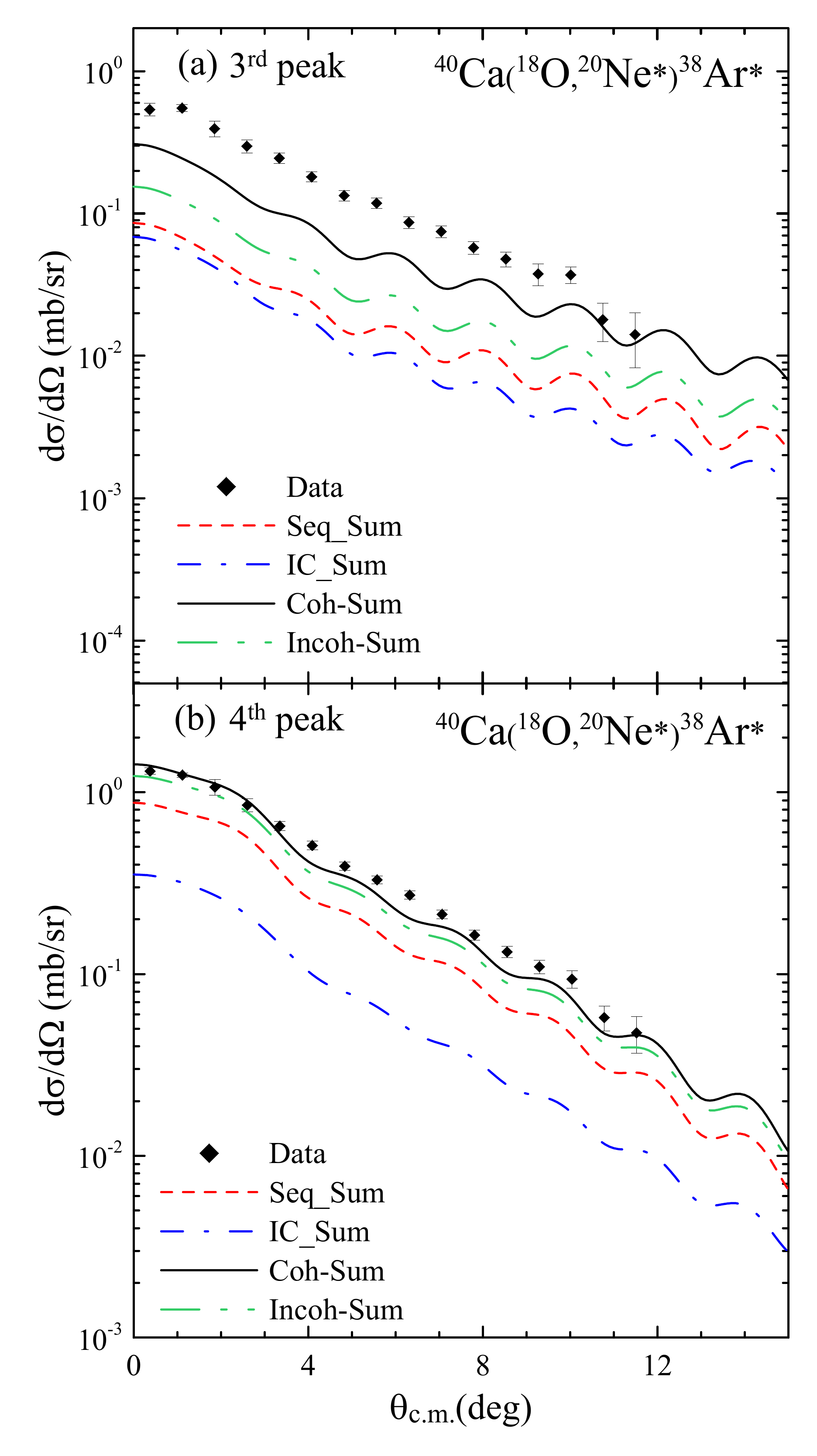} 
\caption {(color online) Comparison between the theoretical and experimental two-proton transfer angular distribution. (a) the angular distribution related to the contribution of the unresolved excited states of the third peak in Fig.~\ref{exp-fig1}. (b) contribution of the unresolved excited states of the fourth peak in Fig.~\ref{exp-fig1} is shown. The contribution due to the simultaneous (IC) and sequential (Seq) transfer and the coherent (Coh) and incoherent (Incoh) sum of the two mechanims are shown (see text).}
\label{tr-fig3}
\end{figure}

In Figs.~\ref{tr-fig2} and \ref{tr-fig3} the extracted angular distribution, for the different energy regions corresponding to the four mentioned peaks, are plotted. The angular binning, in the center of mass reference frame, is 1.5$^\circ$ in the ground state to ground state transition (Fig.~\ref{tr-fig2}a), due to the low statistics, and 0.7$^\circ$, corresponding to the detector intrinsic angular resolution, for the transitions to the excited states (Figs. \ref{tr-fig2}b and \ref{tr-fig3}).
The measured angular distributions have a quite structureless shape, mainly given as a result of the convolution of different oscillatory patterns due to different multipolarities. The bell-shaped behaviour, centered near the grazing angle and essentially independent of the ${L}$ transfer, is not present here since the incident energy is high enough \cite{satchler}.\\

\section{THEORETICAL ANALYSIS}

\subsection{\label{Theo} Brief Description of the formalism}

In two-proton transfer reactions, the two protons can be simultaneously or sequentially transferred, as discussed in Ref.~\cite{satchler}. Although both processes contribute during the reaction, it is interesting to analyse them in a separate way in order to scrutinize each individual contribution.\\

\begin{figure} [ht!]
\centering
\includegraphics[scale=0.30]{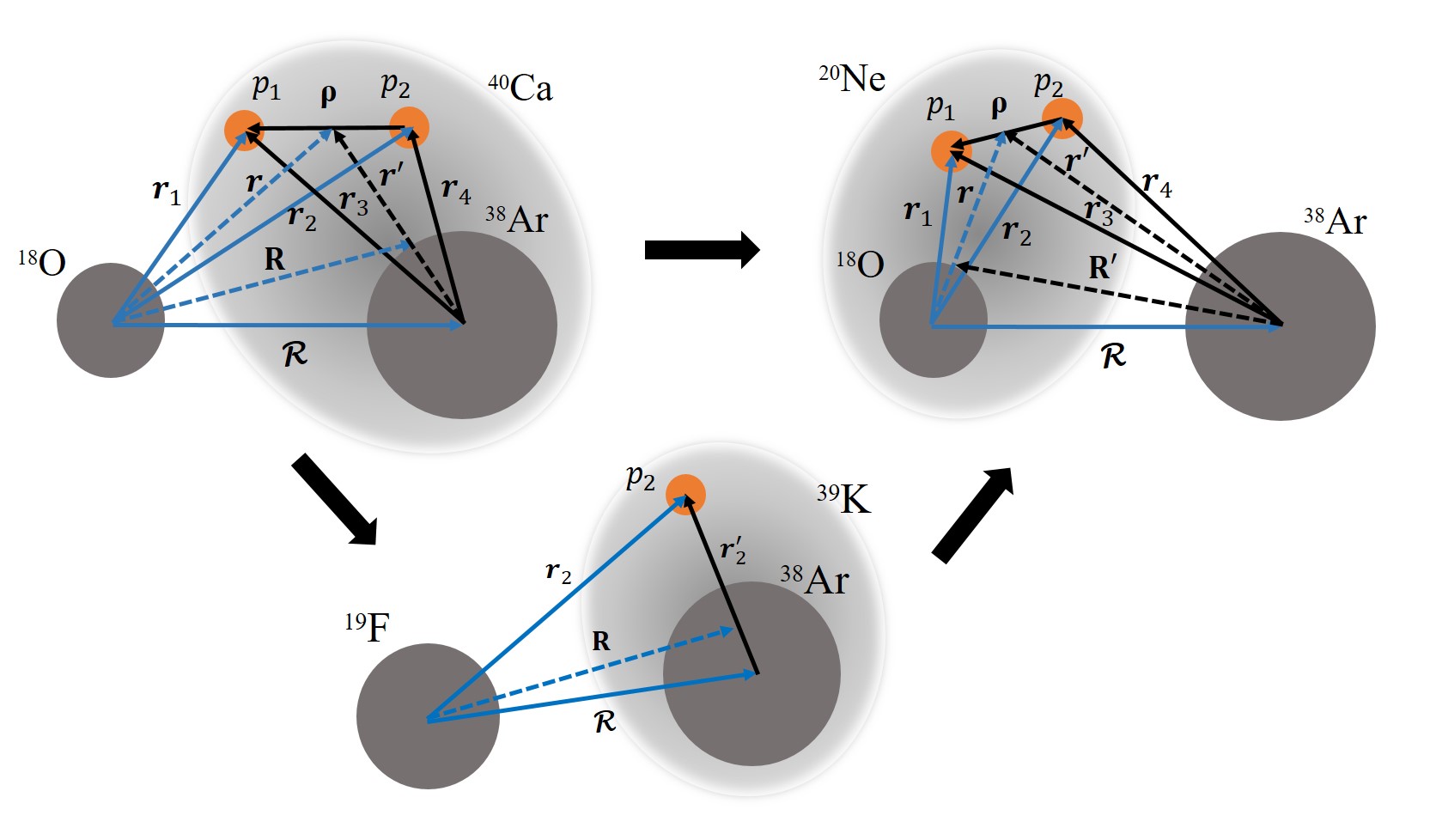} 
\caption {(color online) Coordinates considered in the two-nucleon transfer reaction. The upper path corresponds to the direct two-proton transfer and the lower path is related to the sequential two-proton transfer passing through the intermediate partition.}
\label{tr-fig1}
\end{figure}

Fig.~\ref{tr-fig1} illustrates a two-proton pickup reaction with the coordinates used in the wave functions and interactions. In the figure, the two protons are simultaneously transferred (upper path) or sequentially transferred, passing trough an intermediate partition (lower path).\\ 

For the direct two-proton transfer, the wave function of the initial partition can be written as $\Psi^{(+)}_{\scriptscriptstyle{\alpha}}(\textbf{R},\xi_{i},\xi_{j})$, where $\textbf{R}$ represents the center of mass coordinates between the projectile and target $(a+A)$, while the $\xi_{i}$ are the intrinsic coordinates of the projectile and $\xi_{j}$ the intrinsic coordinates of the target (notice that $\xi_{j} = \lbrace\xi_{j-2},r_{3},r_{4}\rbrace$). $\Psi^{(-)}_{\scriptscriptstyle{\beta}}(\textbf{R}',\xi_{k},\xi_{l})$ is the wave function for the final partition. So, the transfer amplitudes may be determined using

\begin{dmath}[style={\small}]
%\begin{equation}
T^{(direct)}_{\alpha\beta} = \langle \Psi^{(-)}_{\beta}\vert W_{\alpha}\vert\Psi^{(+)}_{\alpha} \rangle ,    
\label{eq-1}
%\end{equation}
\end{dmath}
with $\vert\Psi^{(+)}_{\scriptscriptstyle{\alpha}}\rangle =$ $\sum_{ij}\vert\phi_{a_{i}}\phi_{\scriptscriptstyle{A}_{j}}\chi^{(+)}_{\scriptscriptstyle{\alpha}}\rangle$ $=\sum_{\scriptscriptstyle{\alpha}}\vert\phi_{\scriptscriptstyle{\alpha}}\chi^{(+)}_{\scriptscriptstyle{\alpha}}\rangle$ and $\langle \Psi^{(-)}_{\scriptscriptstyle{\beta}}\vert =$ $\sum_{kl}\langle\phi_{b_{k}}\phi_{\scriptscriptstyle{B}_{l}}\chi^{(-)}_{\scriptscriptstyle{\beta}}\vert =$ $\sum_{\scriptscriptstyle{\beta}}\langle\phi_{\scriptscriptstyle{\beta}}\chi^{(-)}_{\scriptscriptstyle{\beta}}\vert$.\\ 

Above, $\phi_{y}$ (where the sub index $y$ stands for $a_{i},A_{j},b_{k}$ or $B_{l}$) are the intrinsic wave functions of the nuclei in entrance and final partitions with $\chi^{(+)}_{\scriptscriptstyle{\alpha}}$ and $\chi^{(-)}_{\scriptscriptstyle{\beta}}$ being the relative motion wave function, respectively. In this case, $a_{i}, A_{j}, b_{k}$ and $B_{l}$ are all the quantum number needed to determine the state of the $a, A, b,$ and $B$ nuclei. The superscripts $(-)$ and $(+)$ mean the asymptotic ingoing and outgoing wave function of the relative motion, respectively. Finally, the superscript ($direct$) means that the two protons are transferred directly from partition $\alpha $ to $\beta $.\\

The residual interaction $W$, in prior representation, is given by $W_{\alpha} = U(\boldsymbol{\mathscr{R}}) + v(\textbf{r}_{1}) + v(\textbf{r}_{2}) - U(\mathbf{R}) $. The potentials $U(\mathbf{R})$ and $U(\boldsymbol{\mathscr{R}})$ are complex defined to describe the scattering between the $^{18}$O and $^{40}$Ca, as well as, between the $^{20}$Ne and $^{38}$Ar, respectively. So that, the term  $U(\boldsymbol{\mathscr{R}}) - U(\mathbf{R})$ is known as residual remnant potential.   $v(\textbf{r}_{1})$ and $v(\textbf{r}_{2})$ are real potentials which bind each valence nucleon to the core.\\ 
%\textcolor{red}{$W_{\alpha} = U(\mathscr{R}) + v(\mathbf{r}_{1},\mathbf{r}_{2}) - U(R) $}

In this work, the S\~ao Paulo potential was used in the real and imaginary parts of the complex potentials $U(\boldsymbol{\mathscr{R}})$ and $U(\mathbf{R})$ $\left[U(\boldsymbol{x}) = (N_{r} + iN_{i})V_{LE}^{SP}(\boldsymbol{x})\right]$,with $\boldsymbol{x} = \mathbf{R}$ or $\boldsymbol{\mathscr{R}}$. The S\~ao Paulo potential is derived from a double-folding form
\begin{dmath}[style={\small}]
%\begin{equation}
V_{F} = \int\rho_{1}(\mathbf{r}_{1})\mathscr{V}(\textbf{R}-\textbf{r}_{1}+\textbf{r}_{2})\rho_{2}(\mathbf{r}_{2})d\textbf{r}_{1}d\textbf{r}_{2} ,
%\end{equation}
\end{dmath}
being $\rho_{1}$ and $\rho_{2}$, the matter densities of the colliding nuclei and $\mathscr{V}(\textbf{R}-\textbf{r}_{1}+\textbf{r}_{2})$ is the known nucleon-nucleon M3Y interaction \cite{SL1979,FS1985,KOO1996}. When the range of the effective nucleon-nucleon interaction is negligible in comparison with the diffuseness of  the  nuclear  densities, the usual M3Y nucleon-nucleon interaction becomes $V_0 \delta ({\bf R} - {\bf r_1}+ {\bf r_2})$ (zero-range  approach), with $V_0=-456$ MeV$\cdot$fm$^{3}$. The matter densities are determined by considering a two-parameter Fermi-Dirac distribution with radius $R_{0} = 1.31A^{1/3}-0.81$ fm and matter diffuseness $a = 0.56$ fm \cite{078-2,CCG02}, where $A$ is the number of nucleons in the nucleus. This parameterization is usually known as S\~ao Paulo potential systematic. In its local equivalent version, the S\~ao Paulo potential is given by $V^{\scriptscriptstyle{\mathrm{SPP}}}_{\scriptscriptstyle{\mathrm{LE}}}(\mathbf{R},E) = V_{\scriptscriptstyle{\mathrm{F}}}(\mathbf{R})e^{4v^{2}/c^{2}}$ \cite{spp,06,078-1,078-2}, where $v$ is the local relative velocity between the partner nuclei of the collision and $c$ the speed of light.\\  

The intrinsic wave functions for the nucleus composed by a core plus two valence particles are written as

\begin{dmath}[style={\small}]
%\begin{eqnarray}
\phi_{\scriptscriptstyle{IM}_{I}}(\xi_{c},\textbf{r}_{1},\textbf{r}_{2}) =  \sum_{\substack{I_{c},j_{12}, j_{1}\\ j_{2},l_{1},l_{2}}} \mathscr{A}_{j_{1}j_{2}j_{12}}^{I_{c}j_{12}I} \left[\phi_{\scriptscriptstyle{I}_{c}}(\xi_{c}) \otimes \varphi_{\scriptscriptstyle{j}_{12}}(\textbf{r}_{12}) \right]_{\scriptscriptstyle{IM}_{I}}, 
 \label{eq-2}
%\end{eqnarray}
\end{dmath}
with $\varphi_{\scriptscriptstyle{j}_{12}}(\textbf{r}_{12})$ being the two-particle wave function defined by
\begin{dmath}[style={\small}]
\varphi_{\scriptscriptstyle{j}_{12}}(\textbf{r}_{12}) =  \left[\varphi_{\scriptscriptstyle{j}_{1}}(\textbf{r}_{1}) \otimes \varphi_{\scriptscriptstyle{j}_{2}}(\textbf{r}_{2}) \right]_{j_{12}}, 
 \label{eq-2}
%\end{eqnarray}
\end{dmath}

where $\varphi_{\scriptscriptstyle{j}_{1}}(\textbf{r}_{1})$ and $\varphi_{\scriptscriptstyle{j}_{2}}(\textbf{r}_{2})$ are the single-particle wave functions; $\mathscr{A}_{j_{1}j_{2}j_{12}}^{I_{c}j_{12}I}$ stands for the spectroscopic amplitudes of the two valence particles in the single orbits characterized by the total angular momentum $j_{1}$ and $j_{2}$, as well as of the core in the state with total angular momentum $I_{c}$. The total angular momentum of the nucleus is obtained by adding the core spin and the angular momentum resulting by the sum of the total angular momentum of each valence particle. $\phi_{\scriptscriptstyle{I}_{c}}$ is the core wave function. In expression (\ref{eq-2}), $\bm{j}_i = \bm{l}_i+\bm{s}_i$ ($i=1,2$) where $j_{i}$, $l_{i}$ and $s_{i}$ stand for the orbital angular momentum of the single-particle motion, spin, and total angular momentum, respectively.\\

The single-particle wave functions are generated using Woods-Saxon potentials with radii given by $R = r_{0} A_{i}^{1/3}$($A_i$ represents the mass number of the core nucleus in which the valence particle is bound). A reduced radius $r_0 = 1.26$ fm and diffuseness $a = 0.70$ fm were used to generate the single-particle wave functions for the lighter nuclei, whereas 
%\st{$r_{0} = 1.25$} 
$r_{0} = 1.20$ fm and 
%\st{$a = 0.65$} 
$a = 0.60$ fm were used for the heavier nuclei. The depth of these potentials was optimized in order to fit the experimental one-proton binding energy.\\ 

For the sequential two-proton transfer process, the protons are transferred one by one passing through the intermediate partition, for instance
$^{18}$O+$^{40}$Ca ($^{39}$K+p) $\rightarrow ^{19}$F+$^{39}$K ($^{38}$Ar+p) $\rightarrow ^{20}$Ne+$^{38}$Ar, so that the wave function of the nucleus composed by core and valence particle is given by
%$a+A(B+2n) \rightarrow c(a+n)+C(B+n) \rightarrow b(a+2n)+B$
\begin{dmath}[style={\small}]
%\begin{eqnarray}
\phi_{\scriptscriptstyle{IM}_{I}}(\xi_{c},\textbf{r}) =  \sum_{I_{c}lj} \mathscr{A}_{lsj}^{jI_{c}I} \left[\phi_{\scriptscriptstyle{I}_{c}\scriptscriptstyle{M}_{\scriptscriptstyle{I}_{c}}}(\xi_{c}) \otimes \varphi_{\scriptscriptstyle{jm}}(\textbf{r}) \right]_{\scriptscriptstyle{IM}_{\scriptscriptstyle{I}}}. 
 \label{eq-3}
%\end{eqnarray}
\end{dmath}

The transfer amplitude corresponding to the sequential two-particle transfer can be obtained, in prior-prior representation, following Ref.~\cite{satchler}
\begin{dmath}[style={\small}]
%\begin{equation}
T^{(seq)}_{\alpha\beta} = \sum_{\gamma}\langle \Psi^{(-)}_{\beta}\vert  W_{\gamma}\vert\phi_{\gamma}\rangle \tilde{G}^{(+)}_{\gamma}\langle\phi_{\gamma}\vert W_{\alpha}\vert\Psi^{(+)}_{\alpha} \rangle - \langle\Psi^{(-)}_{\beta}\vert  \phi_{\gamma}\rangle\langle\phi_{\gamma}\vert W_{\alpha}\vert\Psi^{(+)}_{\alpha} \rangle  
\label{eq-T-seq}
\end{dmath}

In expression (\ref{eq-T-seq}), $\tilde{G}^{(+)}_{\gamma}$ is the distorted-wave Green function, in the $\gamma$ partition, represented by $\tilde{G}^{(+)}_{\gamma} = \left[E-\epsilon_{\gamma}-K_{\gamma} - \langle \phi_{\gamma} \vert V_{\gamma}\vert \phi_{\gamma} \rangle\right]^{-1}$, where $\epsilon_{\gamma}$ and $K_{\gamma}$ are the intrinsic energy and kinetic energy operator in that partition, respectively. Besides, the second term in expression (\ref{eq-T-seq}) corresponds to the non-orthogonality correction and $\gamma$ refers to the channels considered in the intermediate partition.\\ 

In fact, in a full quantum treatment of the transfer process both direct and sequential transfer amplitudes should be considered in the same calculation, so that the transfer amplitudes to consider should be given by $T_{\alpha\beta} = \vert T^{(direct)}_{\alpha\beta} + T^{(seq)}_{\alpha\beta} \vert$,
which includes the non-orthogonal term deriving from the limited model space of both the direct and sequential calculations. 

However, the second-order calculations recently applied to calculate the cross sections for the $(p,t)$ reactions, for example in Ref. \cite{Pot11}, are still not sufficiently developed to account for the inelastic excitation of the involved nuclei, which are relevant routes when considering heavy-ion-induced reactions \cite{Bon77, Lem77, Per12}.
Our approach, already adopted in Refs. \cite{CCB13, ECL16, CLL17, CFC17, PSV17, CLL18}, is thus to perform the one-step and two-step calculations separately.

\subsection{Shell model calculations}

The one- and two-proton spectroscopic amplitudes for the projectile and target overlaps were derived from shell-model calculations using the NuShellX code \cite{nushell}.\\ 
 
For the projectile overlaps, the amplitudes were calculated considering the Zuker-Buck-McGrory (ZBM) effective interaction \cite{zbm}, in which the $^{12}$C nucleus is considered as a closed core with the 1p$_{1/2}$, 1d$_{5/2}$, and 2s$_{1/2}$ as valence orbits for the neutrons and protons. This realistic interaction successfully describes the structure characteristics of the lowest states of the $^{15,16,17}$O isotopes. Recently, this interaction has been used to derive the one- and two-neutron spectroscopic amplitudes for the overlaps involving the $^{16,17,18}$O and $^{13,14,15}$C isotopes in experiments where a beam of $^{18}$O bombarded the targets $^{12,13}$C~\cite{CCB13,CFC17}, $^{16}$O~\cite{ECL16,CLL17,LEL18}, $^{28}$Si~\cite{CLL18} and $^{64}$Ni~\cite{PSV17}. The experimental angular distributions for the one- and two-neutron stripping transfer reactions were  described quite well. \\

\begin{table}[ht!]
\begingroup
\caption{Comparison between the $^{38}$Ar, $^{39}$K and $^{40}$Ca experimental spectra and the one obtained by shell model calculation considering the ZBM2-modified interaction.} 
\setlength{\tabcolsep}{2pt} % Default value: 6pt
\renewcommand{\arraystretch}{1.3} % Default value: 1
\centering
%\begin{scriptsize}
\begin{small}
\begin{tabular}{ccc}
\hline\hline 
\multicolumn{3}{c}{$^{40}\mathrm{Ca}$} \\ \hline
$I^{\pi}$     & E$_{\mathrm{Exp.}}$(MeV)  & E$_{\mathrm{Theo.}}$(MeV)  \\ \hline
$0_{1}^{+}$ & 0 & 0 \\
$0_{2}^{+}$ & 3.353 & 3.538  \\
$3_{1}^{-}$ & 3.737 & 4.614  \\
$2_{1}^{+}$ & 3.904 & 4.117  \\ \hline \hline
\multicolumn{3}{c}{$^{39}\mathrm{K}$} \\ \hline
$I^{\pi}$     & E$_{\mathrm{Exp.}}$(MeV)  & E$_{\mathrm{Theo.}}$(MeV) \\ \hline
$3/2_{1}^{+}$ & 0 & 0  \\
$1/2_{1}^{+}$ & 2.523 & 1.998  \\
$7/2_{1}^{-}$ & 2.814 & 2.119  \\
$3/2_{1}^{-}$ & 3.019 & 3.196  \\
$9/2_{1}^{-}$ & 3.597 & 3.544  \\
$5/2_{1}^{-}$ & 3.883 & 4.106  \\
$3/2_{2}^{+}$ & 3.939 & 4.469  \\
$11/2_{1}^{-}$ & 3.944 & 3.314 \\
$3/2_{2}^{-}$ & 4.082 & 4.363 \\
$1/2_{2}^{+}$ & 4.096 & 4.718  \\
$7/2_{2}^{-}$ & 4.127 & 3.935  \\

\hline \hline
\multicolumn{3}{c}{$^{38}\mathrm{Ar}$} \\ \hline
$I^{\pi}$     & E$_{\mathrm{Exp.}}$(MeV)  & E$_{\mathrm{Theo.}}$(MeV)  \\ \hline
$0_{1}^{+}$ & 0 & 0 \\
$2_{1}^{+}$ & 2.168 & 2.201  \\
$0_{2}^{+}$ & 3.378 & 3.862  \\
$3_{1}^{-}$ & 3.810 & 3.135  \\
$2_{2}^{+}$ & 3.936 & 3.418  \\
$2_{3}^{+}$ & 4.565 & 4.328  \\
$5_{1}^{-}$ & 4.586 & 3.731  \\
$3_{2}^{-}$ & 4.877 & 4.782  \\
$2_{4}^{+}$ & 5.157 & 4.813  \\
$4_{1}^{+}$ & 5.349 & 4.405  \\
$3_{2}^{-}$ & 5.513 & 5.224  \\
$2_{5}^{+}$ & 5.595 & 5.340  \\
$5_{2}^{-}$ & 5.659 & 5.271  \\
$3_{3}^{-}$ & 5.825 & 5.631  \\
$4_{2}^{+}$ & 6.053 & 5.420  \\
$2_{6}^{+}$ & 6.250 & 5.560  \\
$4_{3}^{+}$ & 6.276 & 6.071  \\
$6_{1}^{+}$ & 6.409 & 5.120  \\
$2_{7}^{+}$ & 6.520 & 6.184  \\
$5_{3}^{-}$ & 6.674 & 6.227  \\
$6_{2}^{+}$ & 7.289 & 6.355  \\ \hline \hline
 
\end{tabular}
%\end{scriptsize}
\end{small}
\label{tr-spectra}
\endgroup
\end{table}

\begin{table}[ht!]
\begingroup
\caption{Comparison between the experimental and theoretical predictions of the reduced electric quadrupole (B(E2)) and octupole (B(E3)) transition probabilities for the  $^{38}$Ar, $^{39}$K and $^{40}$Ca nuclei. \textcolor{blue}{$^{(a)}$}\cite{raman}; \textcolor{blue}{$^{(b)}$}\cite{kibedi};
\textcolor{blue}{$^{(c)}$}\cite{EL1973};
\textcolor{blue}{$^{(d)}$}\cite{EG1969};
\textcolor{blue}{$^{(e)}$}\cite{FLS1996};
\textcolor{blue}{$^{f}$}\cite{RJP1968}}. 
% f -> d   g -> e    h -> f
%\textcolor{blue}{$^{(d)}$}\cite{KOW1976};
%\textcolor{blue}{$^{(e)}$}\cite{KSL1976};

\setlength{\tabcolsep}{2pt} % Default value: 6pt
\renewcommand{\arraystretch}{1.3} % Default value: 1
\centering
%\begin{scriptsize}
\begin{small}
\begin{tabular}{cccc}
\hline\hline 
\multicolumn{4}{c}{$^{40}\mathrm{Ca}$} \\ \hline
 %     & $I^{\pi}$ & Exp. & Theo. \\ \hline
B(E2)$(e^{2}fm^{4})$
  & $I_{i}^{\pi} \rightarrow I_{f}^{\pi}$ & Exp. & Theo. \\
             & $0_{1}^{+} \rightarrow 2_{1}^{+}$ & $99^{\textcolor{blue}{(a)}} $  & 105 \\
B(E3)$(e^{2}fm^{6})$       & $I_{i}^{\pi} \rightarrow I_{f}^{\pi}$ & Exp. & Theo. \\
             & $0_{1}^{+} \rightarrow 3_{1}^{-}$ & 11.800\textcolor{blue}{$^{(b)}$} & 11.420 \\
\hline \hline
\multicolumn{4}{c}{$^{39}\mathrm{K}$} \\ \hline

B(E2)$(e^{2}fm^{4})$& $I_{i}^{\pi} \rightarrow I_{f}^{\pi}$ & Exp.  & Theo. \\
                    & $1/2_{1}^{+} \rightarrow 3/2_{1}^{+}$ & 6.9\textcolor{blue}{$^{(f)}$} ; $22$\textcolor{blue}{$^{(c)}$} & 27.4 \\
B(E3)$(e^{2}fm^{6})$& $I_{i}^{\pi} \rightarrow I_{f}^{\pi}$ & Exp. & Theo. \\
                    & $3/2_{1}^{+} \rightarrow 7/2_{1}^{-}$ & 124 \textcolor{blue}{$^{(f)}$} & 106    \\
                    & $3/2_{1}^{+} \rightarrow 3/2_{1}^{-}$ & 269 \textcolor{blue}{$^{(f)}$} & 217 \\
                    & $3/2_{1}^{+} \rightarrow 9/2_{1}^{-}$ & 694 \textcolor{blue}{$^{(f)}$} & 90  \\
                    & $3/2_{1}^{+} \rightarrow 5/2_{1}^{-}$ & 549 \textcolor{blue}{$^{(f)}$} & 1233    \\

\hline \hline
\multicolumn{4}{c}{$^{38}\mathrm{Ar}$} \\ \hline
B(E2)$(e^{2}fm^{4})$& $I_{i}^{\pi} \rightarrow I_{f}^{\pi}$ & Exp.   & Theo.\\
%                    & $0_{1}^{+} \rightarrow 2_{1}^{+}$     & 130\textcolor{blue}{$^{(a)}$};122\textcolor{blue}{$^{(e)}$};$121\pm 7.6$\textcolor{blue}{$^{(f,g)}$} & 228.5\\
                    & $0_{1}^{+} \rightarrow 2_{1}^{+}$     & 130\textcolor{blue}{$^{(a)}$};$121\pm 7.6$\textcolor{blue}{$^{(d,e)}$} & 228.5\\
                    & $0_{2}^{+} \rightarrow 2_{1}^{+}$     &  $10.63\pm 0.76$\textcolor{blue}{$^{(e)}$}  & 15.29\\
%                    & $0_{1}^{+} \rightarrow 2_{2}^{+}$     & 64\textcolor{blue}{$^{(e)}$};42\textcolor{blue}{$^{(f)}$}  & 11.43 \\
                    & $0_{1}^{+} \rightarrow 2_{2}^{+}$     & 42\textcolor{blue}{$^{(d)}$}  & 11.43 \\
%                    & $2_{1}^{+} \rightarrow 2_{2}^{+}$     & 52\textcolor{blue}{$^{(e)}$}; $56\pm 11$\textcolor{blue}{$^{(g)}$}  & 241.9     \\  
                    & $2_{1}^{+} \rightarrow 2_{2}^{+}$     &  $56\pm 11$\textcolor{blue}{$^{(e)}$}  & 241.9     \\  
                    & $4_{1}^{+} \rightarrow 2_{1}^{+}$     & $7.59\pm 2.28$ \textcolor{blue}{$^{(e)}$}   & 13.5\\  
                    & $4_{1}^{+} \rightarrow 2_{2}^{+}$     & $235.3\pm 68.3$\textcolor{blue}{$^{(e)}$}  & 41.1 \\ 
                    & $6_{2}^{+} \rightarrow 4_{1}^{+}$     & $607\pm 304$\textcolor{blue}{$^{(e)}$} & 37.26 \\
%                    & $5_{1}^{-} \rightarrow 3_{1}^{-}$     & 2.20\textcolor{blue}{$^{(d)}$}; $1.44\pm 0.15$\textcolor{blue}{$^{(g)}$}   &  -    \\
                    & $5_{1}^{-} \rightarrow 3_{1}^{-}$     &  $1.44\pm 0.15$\textcolor{blue}{$^{(e)}$}   &  -    \\

                    & $5_{2}^{-} \rightarrow 3_{1}^{-}$     &  $22\pm 5.3$\textcolor{blue}{$^{(e)}$}   &  -    \\

B(E3)$(e^{2}fm^{6})$& $I_{i}^{\pi} \rightarrow I_{f}^{\pi}$ & Exp. & Theo.\\
                    & $0_{1}^{+} \rightarrow 3_{1}^{-}$     & 9500\textcolor{blue}{$^{(b)}$} & 1251 \\
\hline\hline

\end{tabular}
%\end{scriptsize}
\end{small}
\label{tr-transition}
\endgroup
\end{table}

As regards the target overlaps, to properly describe the structure of the $^{38}$Ar$_{g.s.}$, which has two holes in the $sd$ shell, the full $sd$-$pf$ shells should be considered. However, this calculation requires the use of a powerful computing. 
An approach to skip the computational difficulty of performing a large-scale shell model calculation in the $sd$-$pf$ shells is to control the number of nucleons promoted from the $sd$ shell to the $pf$ one. This procedure has been adopted in Refs.~\cite{RPB2002,SMJ2000}. 
Our approach was to consider a reduced model space without any other additional constraints. 
In this sense, the model space composed by the 2s$_{1/2}$, 1d$_{3/2}$, 1f$_{7/2}$, and 2p$_{3/2}$ valence orbits for protons and neutrons was adopted. 
We have considered the phenomenological interaction (named ZBM2-modified) built to describe the Ca isotopes spectra \cite{CLM2001} and modified  to reproduce the $^{38}$K spectrum. Besides, the authors obtained a better description of the difference in mean-square charge radii between the isomer state $^{38}\mathrm{K}(0^{+})$ and the ground state $^{38}\mathrm{K}(3^{+})$~\cite{BPN2014}. This interaction is a modified version of the one used in Ref.~\cite{RCN1997}. The two-body matrix elements for the particles in the $sd$ shell were taken from the Windenthal interaction \cite{W1984}. The Kuo-Brown interaction was used for the particles in the $pf$ shell \cite{PZ1981}, and the cross shell interaction was taken from Ref.~\cite{KLS1969}. The single-particle energies have been considered to reproduce the spectrum of the $^{29}$Si nucleus. 
\begin{comment}
\st{the effective shell-model Hamiltonian has been obtained from a residual surface delta interaction (SDI)}\cite{vpth,vpnp}
\end{comment} 
As one clearly can realize, the orbits up to the 1d$_{5/2}$ are completely filled in this model space. Moreover, no one proton or neutron is promoted to the 1f$_{5/2}$ and 2p$_{1/2}$ orbits.\\

Table~\ref{tr-spectra} shows a comparison of the theoretical results obtained for the $^{38}$Ar, $^{39}$K, and $^{40}$Ca energy spectra with the experimental values. The comparison between the experimental and theoretical predictions of the reduced electric quadrupole and octupole transition probabilities for these nuclei can be seen in Table~\ref{tr-transition}. One observes a reasonable good agreement.
%The description of the energies of the $3_{1}^{-}$ and $2_{1}^{+}$ states of the $^{40}$Ca does not have a good description since the discrepancy for the $2_{1}^{+}$ state, for instance, is around 2 MeV. 

\subsection{Reaction calculations}

%The predicted and experimental low-lying spectra for the heavier nuclei involved in the two-proton transfer reaction $^{40}$Ca($^{18}$O,$^{20}$Ne)$^{38}$Ar are in Table XX, where a good agreement may be observed.
We performed calculations for the $^{40}\mathrm{Ca}(^{18}\mathrm{O},$ $^{20}\mathrm{Ne})^{38}\mathrm{Ar}$ two-proton transfer angular distributions considering the finite-range coupled reaction channels (CRC) and coupled channels Born approximation (CCBA) approaches, using the FRESCO code \cite{fresco}, \cite{fresco2}. In the present calculations, the S\~ao Paulo double-folding potential \cite{spp} was used in the real and imaginary parts of the optical potential  for the ingoing, intermediate, and outgoing partitions. 
The same potential has been used to describe elastic and inelastic scattering data at the same energy in ref. \cite{Cavallaro_frontiers}.
In the initial partition, the imaginary part was multiplied by a normalization coefficient equal to 0.6 to account for all the channels not explicitly included in the system of coupled equations, like fusion and coupling to bound (with high excitation energy) and continuum states \cite{06}. 
Usually, in the intermediate and outgoing partitions, the strength coefficient of the imaginary part is set to 0.78 \cite{078-1, 078-2}, when no couplings are considered between the states of nuclei in that partition. However, when the couplings between the ground and inelastic states are explicitly included in the partition, as in the present case, the strength coefficient of the imaginary part of the final partition is also set to 0.6. This is the typical adopted prescription \cite{06, Per12, Oliveira2013, CCB13, ECL16, CLL17, CFC17, PSV17, CLL18, LEL18}. 
A check of the sensitivity of the calculations to changes of the strength coefficient of the imaginary part has been performed showing that the transfer cross section is not significantly affected by such changes. \\              

The two-proton transfer reaction was analyzed in two different ways. 
Firstly, we assumed that both valence protons are simultaneously transferred, as correlated particles. The independent coordinates scheme was considered to carry out this direct transfer calculation in the CRC approach. Then, the coordinates of the two valence protons ($r_1$, $r_2$, $r_3$, $r_4$ in Fig.~\ref{tr-fig1}) are transformed into the coordinate of the center of mass of the system composed by the two protons, and the coordinates of the relative motion of them (represented by the coordinates $\rho$ and $r$ in Fig.~\ref{tr-fig1}). This canonical coordinate transformation is known as the Talmi-Moshinsky transformation \cite{moshinsky}. 
Secondly, we assume that the two protons are transferred one by one passing through the intermediate partition $^{19}$F+$^{39}$K. This sequential two-proton transfer process was performed considering the CCBA. In this approach we couple excited states to the ground state to infinite order (coupled channels) in the initial and final partitions, and the couplings among the partitions to first order.\\ 

\begin{figure} [ht!]
\centering
\includegraphics[scale=0.25]{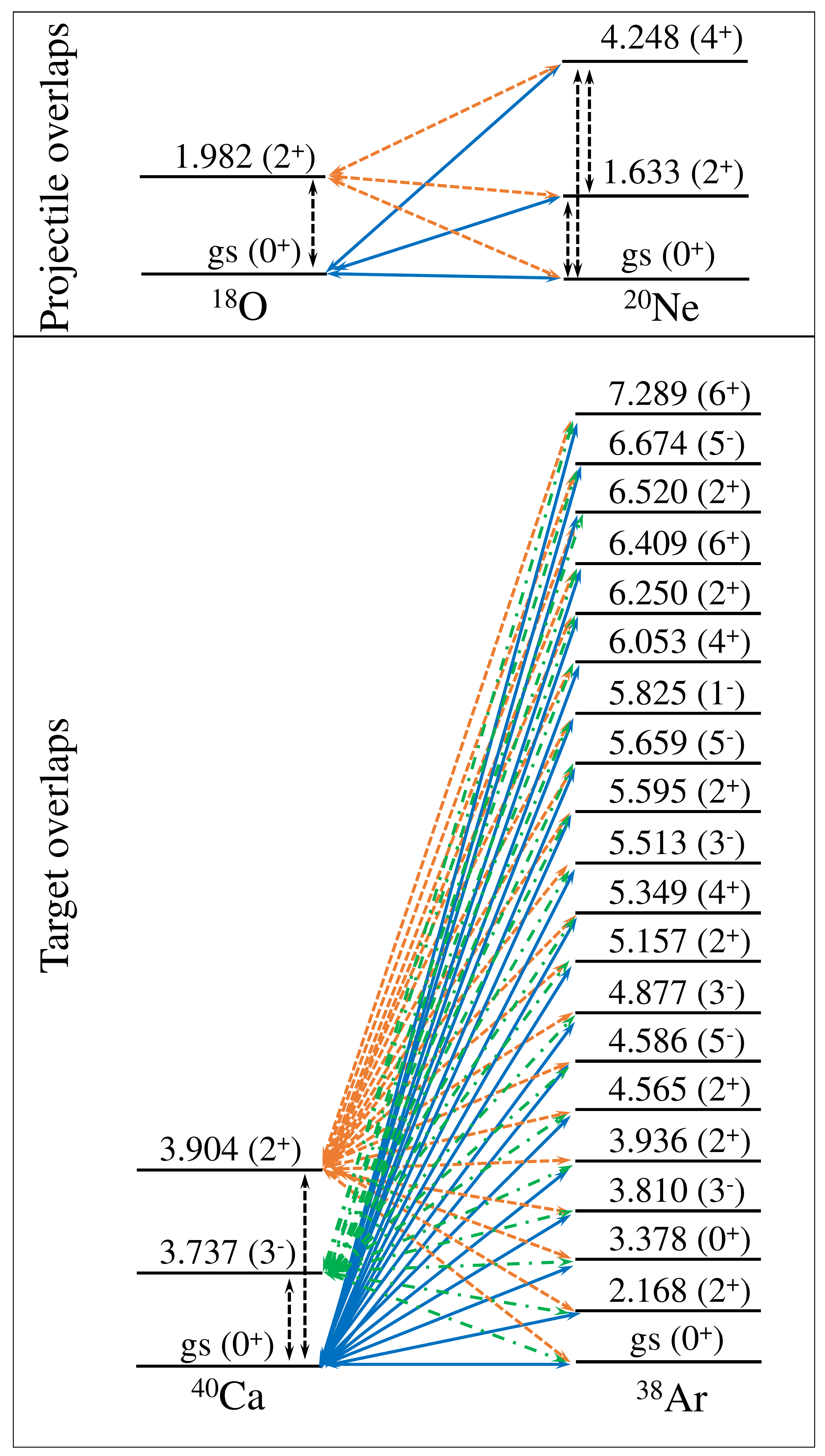} 
\caption {(color online) Coupling scheme considered in the two-proton direct transfer calculations using the independent coordinates scheme.}
\label{tr-fig1a}
\end{figure}

\begin{figure} [ht!]
\centering
\includegraphics[scale=0.2]{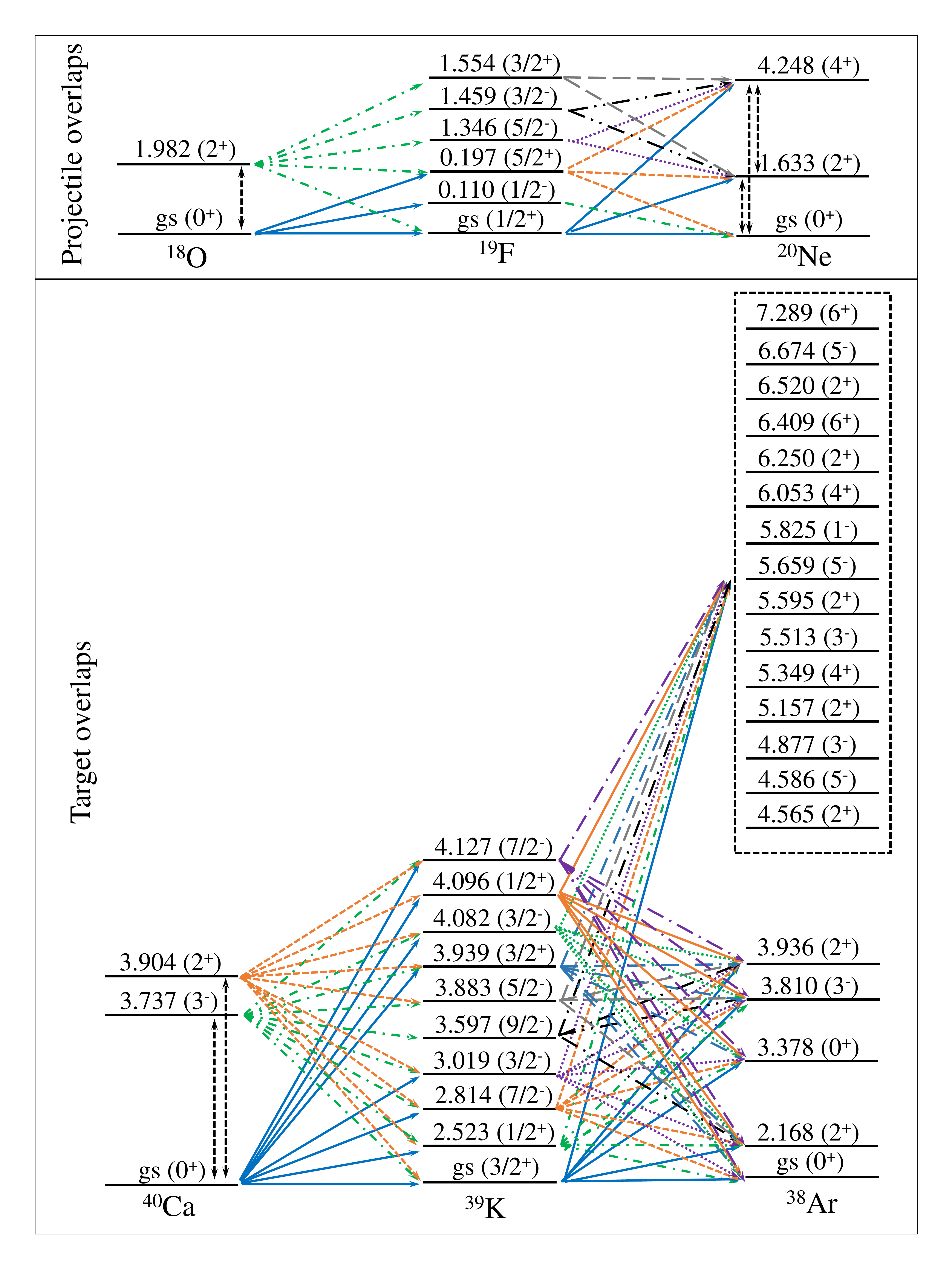} 
\caption {(color online) Coupling scheme considered in the two-proton sequential transfer calculations. }
\label{tr-fig1b}
\end{figure}

The coupling schemes considered in the two-proton transfer calculations are shown in Fig.~\ref{tr-fig1a} for the simultaneous transfer and Fig.~\ref{tr-fig1b} for the sequential transfer. The one phonon quadrupole state of both projectile and target was accessed considering the rotational model. Moreover, in order to derive the reduced electric quadrupole transition probabilities $B(E_{2};0^{+}\rightarrow 2^{+})$ and nuclear deformation length $\delta_{2}$, the deformation parameters $\beta_{2}=0.355$ and $\beta_{2}=0.123$ were used \cite{raman} for the projectile and target, respectively. The reduced electric octupole transition probability $B(E_{3};0^{+}\rightarrow 3^{-})$ and the octupole nuclear deformation $\delta_{3}$ for the target were obtained from the deformation parameter $\beta_{3} = 0.33$ \cite{kibedi}. 
The one- and two-proton spectroscopic amplitudes for the projectile and target overlaps are listed in the tables of appendix \ref{ap1} and \ref{ap2}.\\

\section{\label{results} RESULTS AND DISCUSSION}

In Figs.~\ref{tr-fig2} and \ref{tr-fig3} we show the theoretical angular distributions obtained for two-proton transfer in comparison with the experimental data. All the theoretical curves are convoluted by the experimental angular resolution.\\ 

In Fig.~\ref{tr-fig2}a the measured data and calculations correspond to the channel where ejectile and residual nucleus are in the ground state (first peak of Fig.~\ref{exp-fig1}). From the theoretical results, one can observe that both direct (IC) and sequential (Seq) two-proton transfer processes are of the same order of magnitude and contribute to well describe the cross section. This suggests an existing degree of correlation between the transferred protons.\\

In Fig.~\ref{tr-fig2}b (second peak of Fig.~\ref{exp-fig1}) the experimental angular distribution corresponding to the transition to the sum of the $^{20}\mathrm{Ne}_{1.63}(2^{+})+^{38}\mathrm{Ar}_{\mathrm{gs}}(0^{+})$ and $^{20}\mathrm{Ne}_{\mathrm{gs}}(0^{+})+^{38}\mathrm{Ar}_{2.17}(2^{+})$ channels is shown. The theoretical angular distributions corresponding to the sum of the two channels populated by direct (IC$_{-}$Sum) and sequential (Seq$_{-}$Sum) processes are compared with the experimental data.
One can observe that the sequential and direct two-proton transfer mechanisms compete at very forward angles, but as the angle increases the sequential process becomes dominant.\\  

 The cross sections integrated in the angular range $0^{\circ}\leq \theta_{\mathrm{c.m.}}\leq 12^{\circ}$ for the angular distributions of Fig.~\ref{tr-fig2} are listed in Table~\ref{table-crosssection-1}. As one can see, the channel in which $^{20}$Ne is in its $2^{+}$ first excited state has the strongest cross section, as consequence of the larger quadrupole deformation of $^{20}$Ne compared to $^{38}$Ar.\\

\begin{table}
\begingroup
\caption{Integrated cross sections in the angular range $0^{\circ}\leq \theta_{\mathrm{c.m.}}\leq 12^{\circ}$ for each channel that might contribute to the experimental cross section calculated by direct (IC) and sequential (Seq) mechanism (see text).} 
\setlength{\tabcolsep}{6pt} % Default value: 6pt
\renewcommand{\arraystretch}{1.7} % Default value: 1
\centering
%\begin{scriptsize}
\begin{small}
\begin{tabular}{c|c|c|c|c}
\hline\hline 
%%%%%%%%%%%%%%%%%%%%%%%%%%%%%%%%%%%%%%%%%%%%%%%%%%%%%%%%%%%%%%%%%%%%%%%%%%
%%%%%%%%%%%%%%%%%%%%%%%%%%%%%%%%%%%%%%%%%%%%%%%%%%%%%%%%%%%%%%%%%%%%%%%%%%
 \multicolumn{5}{c}{Channels corresponding to the 1$^{st}$ and 2$^{nd}$ peak (Fig.~\ref{tr-fig2})} \\ \hline\hline

 \multirow{2}{*}{Final Channel}       & \multicolumn{4}{c}{Theoretical cross sections (nb)}   \\ \cline{2-5}
                                      & \multicolumn{2}{c|}{Direct (IC)}  &  \multicolumn{2}{c}{Seq} \\ \hline 

$^{20}\mathrm{Ne}_{\mathrm{g.s}}(0^{+})+^{38}\mathrm{Ar}_{\mathrm{gs}}(0^{+})$ & \multicolumn{2}{c|}{203}& \multicolumn{2}{c}{213} \\ \hline
$^{20}\mathrm{Ne}_{\mathrm{gs}}(0^{+})+^{38}\mathrm{Ar}_{2.17}(2^{+})$         & \multicolumn{2}{c|}{38} & \multicolumn{2}{c}{90} \\ \hline
$^{20}\mathrm{Ne}_{\mathrm{1.63}}(2^{+})+^{38}\mathrm{Ar}_{\mathrm{gs}}(0^{+})$& \multicolumn{2}{c|}{761}  & \multicolumn{2}{c}{844}  \\ \hline
%%%%%%%%%%%%%%%%%%%%%%%%%%%%%%%%%%%%%%%%%%%%%%%%%%%%%%%%%%%%%%%%%%%%%%%%%%
%%%%%%%%%%%%%%%%%%%%%%%%%%%%%%%%%%%%%%%%%%%%%%%%%%%%%%%%%%%%%%%%%%%%%%%%%%
\end{tabular}
\end{small}
%\end{scriptsize}
\label{table-crosssection-1}
\endgroup
\end{table}

As already pointed out, we have treated the two-proton transfer reaction through a sequential or direct process separately. Although these two processes compete with each other, they cannot be separated in the experimental measurement. So, the two-proton transfer cross sections should be obtained by the coherent sum between both mechanisms. 
The relative phase $\phi_{0}$ between the direct and sequential transition amplitudes is extracted by the $\chi^{2}$ search. In Fig.~\ref{tr-fig2} we show the results obtained for the coherent sum between the direct and sequential mechanisms corresponding to those channels discussed so far. In general, one observes an improvement of the agreement between theory and experimental data.
 We also include the curves relative to the incoherent sum to guide our understanding of the role of the interference term in the coherent sum. 
 For instance, in the two-proton transfer angular distribution to the ground state, the interference term has a behavior slightly destructive with $\phi_{0} = -120^{\circ}$. For the angular distribution corresponding to the first peak (Fig.~\ref{tr-fig2}b), a constructive interference ($\phi_{0} = -73^{\circ}$) between the sequential and direct two-proton transfer mechanisms is deduced.\\  

\begin{figure} [h!]
\centering
\includegraphics[scale=0.40]{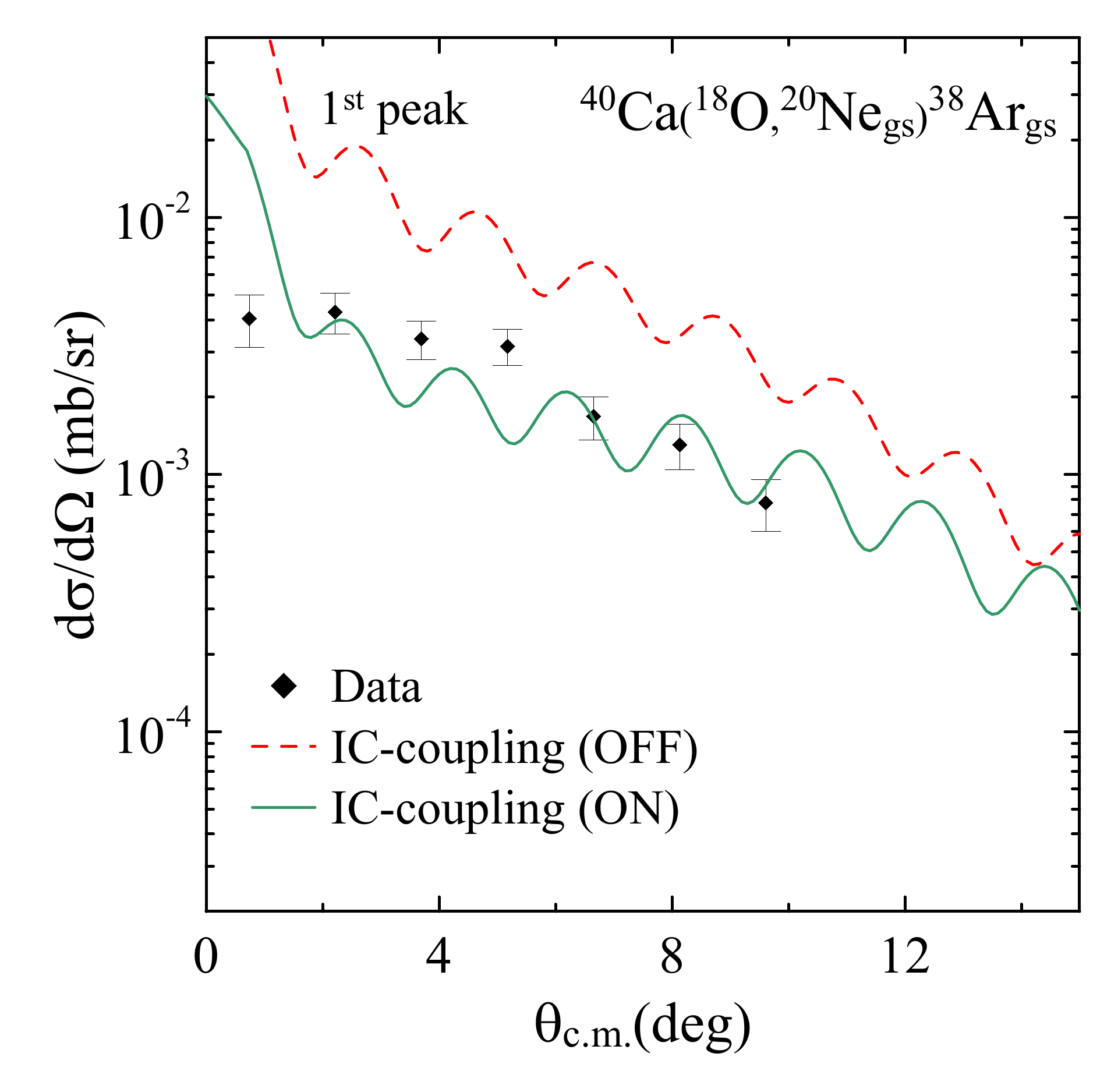} 
\caption {(color online) Comparison between the convoluted theoretical and experimental two-proton transfer angular distribution for the transition to the $^{20}\mathrm{Ne}_{\mathrm{gs}}(0^{+})+^{38}\mathrm{Ar}_{\mathrm{gs}}(0^{+})$ channel. The dashed red curve corresponds to the results in which the couplings between the ground states and inelastic states of the $^{20}$Ne are switched off in the final partition, while in the full green curve these couplings are included. }
\label{tr-fig2a}
\end{figure}

It is important to mention that the coupling between the ground and excited states of the $^{20}$Ne nucleus in the final partition (see Figs.~\ref{tr-fig1a} and \ref{tr-fig1b}) is crucial to describe the order of magnitude of the elastic transfer channel. To illustrate the relevance of the couplings in the final partition, we show in Fig.~\ref{tr-fig2a} the results for the direct two-proton transfer angular distributions where the couplings between the ground and inelastic states of the $^{20}$Ne are switched on and off. The $2_{1}^{+}$ and $4_{1}^{+}$ collective states of the $^{20}$Ne ejectile are accessed, considering the deformation parameter $\beta=0.72$ \cite{raman} in the rotational model frame. A similar behaviour is obtained for the angular distributions of the other measured transitions.\\

The measured angular distribution illustrated in Fig.~\ref{tr-fig3}a is associated with the transition to the channels considered in Table~\ref{table-crosssection-2} (third peak of Fig.~\ref{exp-fig1}). 
From the theoretical integrated cross section, one observes that the most important contribution is given by the $^{20}\mathrm{Ne}_{\mathrm{gs}}(0^{+}) + ^{38}\mathrm{Ar}_{4.57}(2^{+})$ channel followed by the $^{20}\mathrm{Ne}_{\mathrm{gs}}(0^{+}) + ^{38}\mathrm{Ar}_{3.94}(2^{+})$, $^{20}\mathrm{Ne}_{1.63}(2^{+}) + ^{38}\mathrm{Ar}_{2.17}(2^{+})$, and $^{20}\mathrm{Ne}_{4.25}(4^{+}) + ^{38}\mathrm{Ar}_{\mathrm{gs}}(0^{+})$ ones.
Both the sequential and direct two-proton transfer processes have similar integrated cross sections for channels in which the $^{20}$Ne is in its ground state. 
On the other hand, the channels in which the valence protons populated the $2_{1}^{+}$ and $4_{1}^{+}$ excited states of the $^{20}$Ne are preferably populated by the sequential process. 
This agrees with the results obtained for the two-neutron transfer from $^{18}$O to $^{28}$Si~\cite{CLL18} and $^{64}$Ni~\cite{PSV17} nuclei, where the populated states of the residual $^{30}$Si and $^{66}$Ni nuclei are characterize by a large quadrupole deformation as well. In the present case, the states of the $^{20}$Ne ejectile also have strong collectivity. 
The quantum interference concerning the sum of the sequential and direct two-proton transfer amplitudes improves the theoretical predictions, although the experimental angular distribution is slightly underestimated. The obtained $\phi_{0}$ corresponding to the coherent sum between the sequential and direct mechanisms was approximately $0^{\circ} $.\\

 In Fig.~\ref{tr-fig3}b, the angular distribution obtained integrating the energy region corresponding to the fourth peak of Fig.~\ref{exp-fig1} is shown. In this region of the spectrum, the contribution of the population of various excited states is expected. 
 In Table \ref{table-crosssection-2}, the integrated cross section in the angular range $ 0^{\circ} \leq \theta_{\mathrm{c.m.}} \leq 12^{\circ} $  shows the relevance of each channel that can contribute to the experimental cross section. 
 We show the sequential and direct two-proton transfer results, corresponding to the sum of all the theoretical angular distributions associated with the transitions listed in Table \ref{table-crosssection-2}. The strongest channels are  $^{20}\mathrm{Ne}_{\mathrm{1.63}}(2^{+})+^{38}\mathrm{Ar}_{\mathrm{3.94}}(2^{+})$, $^{20}\mathrm{Ne}_{\mathrm{1.63}}(2^{+})+^{38}\mathrm{Ar}_{\mathrm{4.57}}(2^{+})$ and $^{20}\mathrm{Ne}_{\mathrm{1.63}}(2^{+})+^{38}\mathrm{Ar}_{\mathrm{5.16}}(2^{+})$ for which the ejectile nucleus is found in its excited state. Notice, in Table~\ref{table-crosssection-2}, that the sequential mechanism is dominant compared to the direct one, mainly as both valence protons are populating the excited state of the $^{20}$Ne nucleus. The collectivity of these excited states favors the transfer of the two valence protons by a two-step process as pointed out in Refs.~\cite{CLL18, PSV17}.
The interference effect between the sequential and direct mechanisms is slightly constructive ($\phi_{0}$ = $80^{\circ}$), as can be observed in Fig.~\ref{tr-fig3}b. \\    
 
\begin{table}
%\begingroup
\caption{Integrated cross sections in the angular range $0^{\circ}\leq \theta_{\mathrm{c.m.}}\leq 12^{\circ}$ for each channel that might contribute to the experimental cross section calculated by direct (IC) and sequential (Seq) mechanism (see text). For the 4$^th$ peak the cross sections obtained by ZBM2-modified and $vpth$ interaction are listed.} 
\setlength{\tabcolsep}{6pt} % Default value: 6pt
\renewcommand{\arraystretch}{1.7} % Default value: 1
\centering
%\begin{small}
\begin{scriptsize}
\begin{tabular}{c|c|c|c|c}
\hline\hline 
%%%%%%%%%%%%%%%%%%%%%%%%%%%%%%%%%%%%%%%%%%%%%%%%%%%%%%%%%%%%%%%%%%%%%%%%%%
%%%%%%%%%%%%%%%%%%%%%%%%%%%%%%%%%%%%%%%%%%%%%%%%%%%%%%%%%%%%%%%%%%%%%%%%%%
 \multicolumn{5}{c}{Channels corresponding to the 3$^{rd}$ peak (Fig.~\ref{tr-fig3}a)} \\ \hline\hline

 \multirow{2}{*}{Final Channel}       & \multicolumn{4}{c}{Theoretical cross sections (nb)}   \\ \cline{2-5}
%                                      & \multicolumn{2}{c}{ZBM2mod} & \multicolumn{2}{c}{vpth} \\ \cline{2-5}
                                      & \multicolumn{2}{c|}{Direct (IC)}  &  \multicolumn{2}{c}{Seq} \\ \hline 

$^{20}\mathrm{Ne}_{\mathrm{gs}}(0^{+})+^{38}\mathrm{Ar}_{3.38}(0^{+})$         & \multicolumn{2}{c|}{4.85}  & \multicolumn{2}{c}{6.77}  \\ \hline
$^{20}\mathrm{Ne}_{\mathrm{gs}}(0^{+})+^{38}\mathrm{Ar}_{3.81}(3^{-})$         & \multicolumn{2}{c|}{24.11} & \multicolumn{2}{c}{29.37} \\ \hline
$^{20}\mathrm{Ne}_{\mathrm{gs}}(0^{+})+^{38}\mathrm{Ar}_{3.94}(2^{+})$         & \multicolumn{2}{c|}{260.26}& \multicolumn{2}{c}{317.78} \\ \hline
%$^{20}\mathrm{Ne}_{\mathrm{gs}}(0^{+})+^{38}\mathrm{Ar}_{3.38}(0^{+})$           & 3.73   & 6.77      &000   & $2040$ \\ \hline
$^{20}\mathrm{Ne}_{\mathrm{gs}}(0^{+})+^{38}\mathrm{Ar}_{4.57}(2^{+})$         & \multicolumn{2}{c|}{605.31}& \multicolumn{2}{c}{694.32} \\ \hline
$^{20}\mathrm{Ne}_{\mathrm{gs}}(0^{+})+^{38}\mathrm{Ar}_{4.59}(5^{-})$         & \multicolumn{2}{c|}{4.81}  & \multicolumn{2}{c}{9.37}   \\ \hline
$^{20}\mathrm{Ne}_{1.63}(2^{+})+^{38}\mathrm{Ar}_{2.17}(2^{+})$                & \multicolumn{2}{c|}{122.60} & \multicolumn{2}{c}{399.85} \\ \hline 
$^{20}\mathrm{Ne}_{\mathrm{4.25}}(4^{+})+^{38}\mathrm{Ar}_{\mathrm{gs}}(0^{+})$& \multicolumn{2}{c|}{$146.90$}& \multicolumn{2}{c}{$228.53$} \\ \hline \hline
%%%%%%%%%%%%%%%%%%%%%%%%%%%%%%%%%%%%%%%%%%%%%%%%%%%%%%%%%%%%%%%%%%%%%%%%%%
%%%%%%%%%%%%%%%%%%%%%%%%%%%%%%%%%%%%%%%%%%%%%%%%%%%%%%%%%%%%%%%%%%%%%%%%%%
 \multicolumn{5}{c}{Channels corresponding to the 4$^{th}$ peak (Fig.~\ref{tr-fig3}b)} \\  \hline\hline
 \multirow{3}{*}{Final Channel} & \multicolumn{4}{c}{Theoretical cross sections (nb)} \\ %\cline{2-3}
                                & \multicolumn{2}{c}{ZBM2mod} & \multicolumn{2}{c}{vpth} \\ \cline{2-5}
                                & IC  &  Seq. & IC  &  Seq. \\ \hline 
$^{20}\mathrm{Ne}_{\mathrm{gs}}(0^{+})+^{38}\mathrm{Ar}_{5.60}(2^{+})$  &  $171.49$ &  $614.25$ &  $3.09$  &  $1.19$  \\ \hline
$^{20}\mathrm{Ne}_{\mathrm{gs}}(0^{+})+^{38}\mathrm{Ar}_{5.66}(5^{-})$  &  $126.54$ &  $438.29$ &  $44.81$ &  $143.44$ \\ \hline
$^{20}\mathrm{Ne}_{\mathrm{gs}}(0^{+})+^{38}\mathrm{Ar}_{5.83}(3^{-})$  &  $71.34$  &  $262.76$ &  $6.10$  &  $12.53$  \\ \hline
$^{20}\mathrm{Ne}_{\mathrm{gs}}(0^{+})+^{38}\mathrm{Ar}_{6.05}(4^{+})$  &  $0.65$   & $0.88$    &  $0.0082$& $0.11$  \\ \hline
$^{20}\mathrm{Ne}_{\mathrm{gs}}(0^{+})+^{38}\mathrm{Ar}_{6.25}(2^{+})$  &  $81.33$  &  $242.65 $&  $3.95$  &  $6.95 $ \\ \hline
$^{20}\mathrm{Ne}_{\mathrm{gs}}(0^{+})+^{38}\mathrm{Ar}_{6.28}(4^{+})$  &  $0.28$   & $0.29$    &  $0.0402$ & $0.0414$ \\ \hline
$^{20}\mathrm{Ne}_{\mathrm{gs}}(0^{+})+^{38}\mathrm{Ar}_{6.41}(6^{+})$  &  $0.20$   & $2.20$    &  $0.0439$ & $0.0032 $ \\ \hline
$^{20}\mathrm{Ne}_{\mathrm{gs}}(0^{+})+^{38}\mathrm{Ar}_{6.52}(2^{+})$  &  $115.17$ &  $391.42$ &  $1.74$  &  $0.27$  \\ \hline
$^{20}\mathrm{Ne}_{\mathrm{gs}}(0^{+})+^{38}\mathrm{Ar}_{6.67}(5^{-})$  &  $0.21$   &  $4.92$   &  $3.51$  &  $1.32$  \\ \hline
%$^{20}\mathrm{Ne}_{\mathrm{gs}}(0^{+})+^{38}\mathrm{Ar}_{7.29}(6^{+})$  &  $0.08$   & $1.83 \times 10^{-2}$& $1.75 \times 10^{-2}$&   -  \\ \hline

$^{20}\mathrm{Ne}_{\mathrm{1.63}}(2^{+})+^{38}\mathrm{Ar}_{3.94}(2^{+})$& $1283$ & $2630 $ & $7210$ & $7890$  \\  \hline

$^{20}\mathrm{Ne}_{\mathrm{1.63}}(2^{+})+^{38}\mathrm{Ar}_{4.57}(2^{+})$& $3116$  & $8070$  & $28.35$  & $25.03$  \\  \hline
$^{20}\mathrm{Ne}_{\mathrm{1.63}}(2^{+})+^{38}\mathrm{Ar}_{4.59}(5^{-})$& $20.55$ & $58.97$ & $0.98$   & $5.68$  \\  \hline
$^{20}\mathrm{Ne}_{\mathrm{1.63}}(2^{+})+^{38}\mathrm{Ar}_{4.88}(3^{-})$& $178.81$&$456.11$ & $122.31$ &$50.72$  \\  \hline
$^{20}\mathrm{Ne}_{\mathrm{1.63}}(2^{+})+^{38}\mathrm{Ar}_{5.16}(2^{+})$& $1678 $ &  $4670$ & $14.06$  & $4.36$  \\  \hline
$^{20}\mathrm{Ne}_{\mathrm{1.63}}(2^{+})+^{38}\mathrm{Ar}_{5.35}(4^{+})$& $3.62$  & $6.88$  & $0.0015$ & $0.0273$  \\ \hline
$^{20}\mathrm{Ne}_{\mathrm{1.63}}(2^{+})+^{38}\mathrm{Ar}_{5.51}(3^{-})$& $34.68$ & $58.61$ &  14.11    & 7.29   \\ \hline
$^{20}\mathrm{Ne}_{\mathrm{4.25}}(4^{+})+^{38}\mathrm{Ar}_{2.17}(2^{+})$& $47.33$ & $258.80$& $78.57$  & $82.14$  \\ \hline

\end{tabular}
%\end{small}
\end{scriptsize}
\label{table-crosssection-2}
%\endgroup
\end{table}

For the angular distributions considered in Fig.~\ref{tr-fig3}, the model space used in the shell-model calculations might not be enough to properly account for the spectroscopic amplitudes of high-lying excited states of $^{38}$Ar. 
%This could be the reason why some channels have too small cross sections. 
Moreover, the present limited model space for the heavier nuclei does not take into account the $1\mathrm{d}_{5/2}$ orbital, which could produce 2p-4h states with spin $4^{+}$ of higher energies in the $^{38}$Ar \cite{RPB2002}. Therefore, the contribution of the channels $^{20}\mathrm{Ne}_{\mathrm{gs}}(0^{+})+^{38}\mathrm{Ar}_{6.05}(4^{+})$ and $^{20}\mathrm{Ne}_{\mathrm{gs}}(0^{+})+^{38}\mathrm{Ar}_{6.28}(4^{+})$ may still be underestimated. This also could justify the very small cross section obtained in channels in which the $^{38}$Ar nucleus is found in the $6^{+}$  state at 6.41 MeV. 
However, taking into account the constructive interference in the coherent sum of the sequential and direct contributions, the agreement of the sum curve with the experimental data results very good in the case of the fourth peak, while still the theory slightly underestimates the data for the third peak.\\

Additionally, we have calculated the cross section for two-proton transfer using other spectroscopic amplitudes derived from the shell-model calculation, considering a surface delta interaction to obtain the two-body matrix elements \cite{vpth, vpnp}. The results for the cross sections using this interaction are labeled by $vpth$ in Table~\ref{table-crosssection-2}. As one can see, almost all the cross sections derived by using the ZBM2-modified amplitudes are larger than those obtained considering the $vpth$ amplitudes. The results using the $vpth$ amplitudes concentrate all the strength to populate only one $2^{+}$ state of the residual nucleus. Conversely, the cross section derived with the ZBM2-modified amplitudes is spread out to all $2^{+}$ states of $^{38}$Ar, in agreement with the experimental observation.

\section{\label{conclusion} CONCLUSIONS}

The $^{40}$Ca($^{18}$O,$^{20}$Ne)$^{38}$Ar two-proton pickup reaction has been explored at 270 MeV for the first time at forward angles, including zero degrees. Energy spectra and cross section angular distributions for transitions to low-lying states have been extracted. 
The main motivation for this study is related to the interest in a complete study of the reaction mechanism and nuclear structure issues involved in the $^{18}$O+$^{40}$Ca collision, which has been shown an auspicious experimental tool to explore the nuclear matrix elements of the double charge exchange reactions. 

In the analysis we have adopted the finite range coupled reaction channel and coupled channel Born approximation methods to interpret the measured cross sections.
The double-folding São Paulo potential has been used for the ingoing and outgoing partitions.  
Contributions from the direct and sequential transfer mechanism have been both calculated. Since these mechanisms are present in the reaction and can not be experimentally distinguished, the coherent sum has also been performed.\\

The one- and two-proton spectroscopic amplitudes for the projectile and target overlaps have been derived by shell-model calculations.
The structure calculations for the heavier nuclei have been performed considering the reduced model space which includes the 2s$_{1/2}$, 1d$_{3/2}$, 1f$_{7/2}$, and 2p$_{3/2}$ orbitals and the ZBM2-modified effective interaction. The results describe reasonably well the spectrum, as well as the reduced electric quadrupole and octupole transition probabilities. 
The spectroscopic amplitudes corresponding to the overlaps with the $2^{+}$ excited states of $^{38}$Ar spread out the cross section strength among these excited states, in agreement with the experimental observation. 
Some discrepancy between theoretical and experimental values for the energies of the $4_{1}^{+}$, $4_{2}^{+}$, $6_{1}^{+}$ and $6_{2}^{+}$ excited states of $^{38}$Ar is observed. Moreover, the very small theoretical cross section corresponding to those channels might be related to the present limitation in the model space. 
For a better description it would be needed to include at least the 1d$_{5/2}$ orbit in the model space, since the 2p-4h configuration could have a significant contribution to these states. \\           

The cross section angular distribution for the transfer to the ground state channel is well reproduced when the couplings among the ground and inelastic states of $^{20}$Ne are explicitly included in the final partition. 
This channel is populated through the direct or sequential mechanism with very similar strength. On the other hand, the transfer to the excited states proceeds, preferably, by the sequential process, especially in the transitions to the vibrational states of the ejectile. 
In these cases, the high collectivity of such $^{20}$Ne states seems to break the correlation of both transferred protons reducing the direct two-proton transfer contribution to the cross section. Similar behavior was observed in two-neutron transfer reactions, in which both valence neutrons were transferred from $^{18}$O to $^{28}$Si and $^{64}$Ni nuclei and populated the states with high collectivity in the residual nuclei $^{30}$Si and $^{66}$Ni. The main difference in the present calculation is that the two transferred particles are charged.\\

Summarizing, the description of the explored two-proton transfer reaction in terms of spectroscopic amplitudes and differential cross sections for transitions to different populated states is satisfactory.\\

The approach described in the present work
consists in a close cooperation between challenging experimental and theoretical studies, namely 
the measurement of high resolution energy spectra and cross section angular distributions and the comparison with fully microscopic calculations.
This methodology was adopted in the past for two-neutron transfer in lighter systems in similar dynamical conditions also by some of the authors. However, here for the first time its reliability is checked in the two-proton transfer case.
It is possible to conclude that the ($^{18}$O,$^{20}$Ne) reaction can be considered an adequate spectroscopic probe if accompanied by a complete microscopic treatment of reaction and nuclear structure issues. \\

An important impact of this work is the possibility to obtain a complete analysis of double charge exchange reactions in view of their relation with 0$\nu\beta\beta$ decay, for which two-proton transfer reactions could represent the first step of a possible multi-nucleon transfer route that might compete with the direct meson exchange mechanism. Thus the availability of reliable theoretical predictions of such mechanisms, also when experimental data are not available, is a crucial ingredient of this research.

\section*{Acknowledgments}
This project has received funding from the European Research Council (ERC) under the European Union’s Horizon 2020 Research and Innovation Programm (Grant Agreement No. 714625).
The authors acknowledge partial financial support from CNPq, FAPERJ, FAPESP process number 2013/22100-7, 2016/21434-7, CAPES and INCT-FNA (Instituto Nacional de Ci\^ {e}ncia e Tecnologia- F\' isica Nuclear e Aplica\c {c}\~ {o}es).

%\clearpage
%CCCCCCCCCCCCCCCCCCCCCCCCCCCCCCCCCCCCCCCCCCCCCCCCCCCCCCCCCCCCCCCCCCCCCCCCCCCCCCCCCCCCCCCCCCCCCCCCCCCCCCCCC
%    							 		REFERENCIAS
%CCCCCCCCCCCCCCCCCCCCCCCCCCCCCCCCCCCCCCCCCCCCCCCCCCCCCCCCCCCCCCCCCCCCCCCCCCCCCCCCCCCCCCCCCCCCCCCCCCCCCCCCC
\newpage

\clearpage
\appendix
%\chapter{Appendix}
\input{appendix}

\end{document}

%% file: appendix.tex
%%%%%%%%%%%%%%%%%%%%%%%%%%%%%%%%%%%%%%%%%%%%%%%%%%%%%%%%%%%%%%%%%%%%%%%%%%%%%%%%%%%%%%%%%%%%%%%%%%%%%%%%%%%%%%%%%%%
%%%%%%%%%%%%%%%%%%%%%%%%%%%%%%%%%%%%%%%%%%%%%%%%%%%%%%%%%%%%%%%%%%%%%%%%%%%%%%%%%%%%%%%%%%%%%%%%%%%%%%%%%%%%%%%%%%%
%%%%%%%%%%%%%% Appendix %%%%%%%%%%%%%%%%%%%%%%%%%%%%%%%%%%%%%%%%%%%%%%%%%%%%%%%%%%%%%%%%%%%%%%%%%%%%%%%%%%%%%%%%%%%
\section{One-proton spectroscopic amplitudes for the projectile and target overlaps.}
\label{ap1}

\begin{table}[h!]
\begingroup
\LTcapwidth=0.45\textwidth
\caption{One-proton spectroscopic amplitudes concerning the projectile overlaps used in the CCBA calculations for the sequential two-proton transfer reaction, where $n$, $l$ and \textit{$j$} are the principal quantum number, the orbital angular momentum and the spin of the proton orbitals, respectively.} 
\footnotesize
\centering
\setlength{\tabcolsep}{6pt} % Default value: 6pt
\renewcommand{\arraystretch}{1.3} % Default value: 1
\begin{tabular}{|c|c|c|c|}
\hline \textbf{Initial State} & \textbf{$nl_j$}  & \textbf{Final State} & \textbf{S.A.}\\ \hline

\multirow{3}{*}{$^{18}$O$_{g.s.}(0^+)$}
                       & $(2s_{1/2})$ &  $^{19}$F$_{g.s.}(1/2^+)$  & -0.607  \\\cline{2-4} 

                       & $(1p_{1/2})$ &  $^{19}$F$_{0.110}(1/2^-)$  & -0.446  \\\cline{2-4} 
                       
                       & $(1d_{5/2})$ &  $^{19}$F$_{0.197}(5/2^+)$  & -0.644  \\ \hline

\multirow{7}{*}{$^{18}$O$_{1.982}(2^+)$}
                        & $(1d_{5/2})$ &  $^{19}$F$_{g.s.}(1/2^+)$  & -0.596  \\ \cline{2-4}

                        & $(2s_{1/2})$ & \multirow{2}{*}{$^{19}$F$_{0.197}(5/2^+)$} & -0.426 \\ 

                        & $(1d_{5/2})$ &                            & -0.420  \\ \cline{2-4}

                        & $(1p_{1/2})$ &  $^{19}$F$_{1.346}(5/2^-)$ & 0.388  \\ \cline{2-4}

                        & $(1p_{1/2})$ &  $^{19}$F$_{1.459}(3/2^-)$ & -0.402  \\ \cline{2-4} 

                        & $(2s_{1/2})$ & \multirow{2}{*}{$^{19}$F$_{1.554}(3/2^+)$} & -0.668  \\ 

                        & $(1d_{5/2})$ &                            & -0.415  \\ \hline
%%%%%%%%%%%%%%%%%%%%%%%%%%%%%%%%%%%%%%%%%%%%%%%%%%%%%%%%%%%%%%%%%%%%%%%%%%%%%%%%%%%%%%%%%%%%%%%%%%%%%%%
$^{19}$F$_{g.s.}(1/2^+)$                  & $(2s_{1/2})$ & \multirow{3}{*}{$^{20}$Ne$_{g.s.}(0^+)$} & -0.858  \\ 

$^{19}$F$_{0.110}(1/2^-)$                & $(1p_{1/2})$ &                          & 1.270  \\ 

$^{19}$F$_{0.197}(5/2^+)$                 & $(1d_{5/2})$ &                          & -1.174  \\ \hline

$^{19}$F$_{g.s.}(1/2^+)$                  & $(1d_{5/2})$ &\multirow{7}{*}{$^{20}$Ne$_{1.634}(2^+)$} & 0.671  \\ 

\multirow{2}{*}{$^{19}$F$_{0.197}(5/2^+)$}& $(2s_{1/2})$ &                           & 0.692 \\ 

                                          & $(1d_{5/2})$ &                           & 0.642  \\

$^{19}$F$_{1.346}(5/2^-)$                 & $(1p_{1/2})$ &                           & 0.978  \\ 

$^{19}$F$_{1.459}(3/2^-)$                 & $(1p_{1/2})$ &                           & 0.816  \\ 

\multirow{2}{*}{$^{19}$F$_{1.554}(3/2^+)$}& $(2s_{1/2})$ &                           & -0.377  \\ 

                                          & $(1d_{5/2})$ &                          & -0.292  \\ \hline

$^{19}$F$_{0.197}(5/2^+)$                 & $(1d_{5/2})$ & \multirow{2}{*}{$^{20}$Ne$_{4.248}(4^+)$}& -0.646  \\ 

$^{19}$F$_{1.554}(3/2^+)$                 & $(1d_{5/2})$ &                                           & 0.635  \\ \hline

$^{19}$F$_{0.110}(1/2^-)$                 & $(1d_{5/2})$ &  \multirow{7}{*}{$^{20}$Ne$_{4.967}(2^-)$}& 0.073  \\  

$^{19}$F$_{0.197}(5/2^+)$                 & $(1p_{1/2})$ &                                           & 0.034  \\ 

\multirow{2}{*}{$^{19}$F$_{1.346}(5/2^-)$}& $(2s_{1/2})$ &                                           & -0.204  \\ 

                                          & $(1d_{5/2})$ &                                           & 0.639  \\  

\multirow{2}{*}{$^{19}$F$_{1.459}(3/2^-)$}& $(2s_{1/2})$ &                                           & 0.139  \\  

                                          & $(1d_{5/2})$ &                                           & -0.550  \\  

$^{19}$F$_{1.554}(3/2^+)$                 & $(1p_{1/2})$ &                                           & 0.188  \\ \hline

$^{19}$F$_{0.110}(1/2^-)$                 & $(1d_{5/2})$ & \multirow{5}{*}{$^{20}$Ne$_{5.621}(3^-)$}& 0.577  \\  

$^{19}$F$_{0.197}(5/2^+)$  & $(1p_{1/2})$ &                                                         & -0.192  \\  

\multirow{2}{*}{$^{19}$F$_{1.346}(5/2^-)$}& $(2s_{1/2})$ &                                          & 0.285  \\  

                                          & $(1d_{5/2})$ &                                          & 0.110  \\  

$^{19}$F$_{1.459}(3/2^-)$                 & $(1d_{5/2})$ &                                          & -0.170   \\ \hline

$^{19}$F$_{g.s.}(1/2^+)$                  & $(1p_{1/2})$ & \multirow{6}{*}{$^{20}$Ne$_{5.788}(1^-)$}& -0.088  \\ 

$^{19}$F$_{0.110}(1/2^-)$                 & $(2s_{1/2})$ &                                          & 0.095   \\ 

$^{19}$F$_{1.346}(5/2^-)$                 & $(1d_{5/2})$ &                                          & -0.649  \\ 

\multirow{2}{*}{$^{19}$F$_{1.459}(3/2^-)$}& $(2s_{1/2})$ &                                          & 0.114  \\ 

                                          & $(1d_{5/2})$ &                                          & -0.531  \\ 

$^{19}$F$_{1.554}(3/2^+)$                 & $(1p_{1/2})$ &                                          & 0.232  \\ \hline

$^{19}$F$_{g.s.}(1/2^+)$   & $(2s_{1/2})$ & \multirow{3}{*}{$^{20}$Ne$_{6.726}(0^+)$}               & 0.095  \\  

$^{19}$F$_{0.110.}(1/2^-)$ & $(1p_{1/2})$ &                                                         & -0.313  \\  

$^{19}$F$_{0.197}(5/2^+)$  & $(1d_{5/2})$ &                                                         & -0.130  \\ \hline
\end{tabular}
\endgroup
\end{table}

%\clearpage
%%%%%%%%%%%%%%%%%%%%%%%%%%%%%%%%%%%%%%%%%%%%% One-proton spectroscopic amplitudes %%%%%%%%%%%%%%%%%%%%%%%%%%%%%%%%%%%%%%%
%%%%%%%%%%%%%%%%%%%%%%%%%%%%%%%%%%%%%%%%%%%%%%%%%%%%%%%%%%%%%%%%%%%%%%%%%%%%%%%%%%%%%%%%%%%%%%%%%%%%%%%%%%%%%%%%%%%%%%%%%

%\begin{footnotesize}
\begingroup
\setlength{\tabcolsep}{6pt} % Default value: 6pt
\renewcommand{\arraystretch}{1.3} % Default value: 1
%\footnotesize
% \resizebox{0.5\textwidth}{!}{\begin{minipage}{\textwidth}
%\setlength{LTcapwidth}{5.2in}
\LTcapwidth=0.45\textwidth
\begin{longtable}{|c|c|c|c|}
% \begin{minipage}{.5\textwidth}
 \caption{One-proton spectroscopic amplitudes concerning the target overlaps used in the CCBA calculations for the sequential two-proton transfer reaction, where $n$, $l$ and \textit{j} are the principal quantum number, the orbital angular momentum and the spin of the proton orbitals, respectively.} 
\label{tabela-seq-target}\\ 
\hline
% \end{minipage}
%\centering
\multicolumn{1}{|c|}{\textbf{Initial State}} & \multicolumn{1}{c|}{$nl_{j}$} & \multicolumn{1}{c|}{\textbf{Final State}} & \multicolumn{1}{c|}{\textbf{S.A.}} \\ \hline
%\textbf{Initial State} & $nl_{j}$  & \textbf{Final State} & \textbf{S.A.}\\ \hline
\endfirsthead
\multicolumn{4}{c}%
%{{\bfseries \tablename\ \thetable{} -- Continuation of the Table \ref{tabela-seq-target} }} \\ \hline
{{\bfseries  \tablename \thetable{} \ref{tabela-seq-target} continued }} \\ \hline

%\multicolumn{1}{|c|}{\textbf{Initial State}} & \multicolumn{1}{c|}{$nl_{j}$} & \multicolumn{1}{c|}{\textbf{Final State}} & \multicolumn{1}{c|}{\textbf{S.A.}} \\ \hline
\endhead

\endfoot
\endlastfoot
\hline
\multirow{8}{*}{$^{40}$Ca$_{g.s.}(0^+)$}
                                          & $(1d_{3/2})$ &  $^{39}$K$_{g.s.}(3/2^+)$  & 1.787 \\ \cline{2-4}  
                                          & $(2s_{1/2})$ &  $^{39}$K$_{2.523}(1/2^+)$ &-1.277 \\ \cline{2-4}  
                                          & $(1f_{7/2})$ &  $^{39}$K$_{2.814}(7/2^-)$ &-0.718 \\ \cline{2-4}  
                                          & $(2p_{3/2})$ &  $^{39}$K$_{3.019}(3/2^-)$ & 0.223 \\ \cline{2-4}  
                                          & $(1d_{3/2})$ &  $^{39}$K$_{3.939}(3/2^+)$ & 0.218 \\ \cline{2-4}  
                                          & $(2p_{3/2})$ &  $^{39}$K$_{4.082}(3/2^-)$ & 0.172 \\ \cline{2-4}  
                                          & $(2s_{1/2})$ &  $^{39}$K$_{4.096}(1/2^+)$ &-0.362 \\ \cline{2-4}  
                                          & $(1f_{7/2})$ &  $^{39}$K$_{4.127}(7/2^-)$ & 0.245 \\
\end{longtable}
% \end{minipage}}
\endgroup
%\end{footnotesize}

\clearpage
%%%%%%%%%%%%%%%%%%%%%%%%%%%%%%%%%%%%%%%%%%%%%% two-proton spectroscopic amplitudes %%%%%%%%%%%%%%%%%%%%%%%%%%%%%%%%%%%%
%%%%%%%%%%%%%%%%%%%%%%%%%%%%%%%%%%%%%%%%%%%%%%%%%%%%%%%%%%%%%%%%%%%%%%%%%%%%%%%%%%%%%%%%%%%%%%%%%%%%%%%%%%%%%%%%%%%%%%%
\section{Two-proton spectroscopic amplitudes for the projectile and target overlaps.}
\label{ap2}

\begin{table}[h!]
\begingroup
\LTcapwidth=0.45\textwidth
\caption{Two-proton spectroscopic amplitudes concerning the projectile overlaps used in the CRC transfer calculations, where $j_{1}$, $j_{2}$ are the single particle spins and $J$ is the total angular momentum of the transferred protons.} 
\centering
\setlength{\tabcolsep}{2pt} % Default value: 6pt
\renewcommand{\arraystretch}{1.3} % Default value: 1
\begin{tabular}{|c|c|c|c|c|}
\hline \textbf{Initial State} & \textbf{$j_{1}j_{2}$} &   J   & \textbf{Final State} & \textbf{S.A.}\\ \hline

\multirow{12}{*}{$^{18}$O$_{g.s.}(0^+)$}
                       &$1p_{1/2}1p_{1/2}$  &                  &                          & -0.4021  \\ 
                       &$2s_{1/2}2s_{1/2}$  &  0               & $^{22}$Ne$_{g.s.}(0^+)$  &  0.5553  \\ 
                       &$1d_{5/2}1d_{5/2}$  &                  &                          &  0.3681  \\ \cline{2-5}

                       &$1d_{5/2}1d_{5/2}$  &\multirow{2}{*}{2}&\multirow{2}{*}{$^{22}$Ne$_{1.634}(2^+)$}& -0.2728  \\ 
                       &$1d_{5/2}2s_{1/2}$  &                  &                          & -0.4022  \\ \cline{2-5}

                       &$1d_{5/2}1d_{5/2}$  &  4               & $^{22}$Ne$_{4.248}(4^+)$ &  0.2729  \\ \cline{2-5}

                       &$1p_{1/2}1d_{5/2}$  &  2               & $^{22}$Ne$_{4.967}(2^-)$ & -0.0303  \\
\end{tabular}
\endgroup
\end{table}

\begingroup
\renewcommand{\arraystretch}{1.3} % Default value: 1
\begin{scriptsize}
\LTcapwidth=0.45\textwidth
\begin{longtable}{|c|c|c|c|c|}
% \begin{minipage}{.5\textwidth}
\caption{Two-proton spectroscopic amplitudes concerning the target overlaps used in the CRC transfer calculations, where $j_{1}$, $j_{2}$ are the single particle spins and $J$ is the total angular momentum of the transferred protons.} 
\label{tabela-ic-target}\\ \hline
% \end{minipage}%\footnotesize
%\centering
%\multicolumn{4}{|c|}
%\begin{tabular}{|c|c|c|c|}
\textbf{Initial State} & $j_{1}j_{2}$ &   J   & \textbf{Final State} & \textbf{S.A.} \\ \hline
\endfirsthead
\multicolumn{5}{c}%
%{{\bfseries \tablename\ \thetable{} -- Continuation of the Table \ref{tabela-seq-target} }} \\ \hline
{{\bfseries \tablename \thetable{} Table \ref{tabela-ic-target} continued}} \\ \hline
\endhead
%\hline \multicolumn{4}{|r|}{{Continuation of the Table \ref{tabela-seq-target}}} \\ 
%\endfoot
%\endlastfoot
\multirow{14}{*}{$^{40}$Ca$_{g.s.}(0^+)$}
                       &$1d_{3/2}1d_{3/2}$  &\multirow{4}{*}{0}&\multirow{4}{*}{$^{38}$Ar$_{g.s.}(0^+)$} & 0.8249\\ 
                       &$2s_{1/2}2s_{1/2}$  &     &                          & 0.2350 \\ 
                       &$1f_{7/2}1f_{7/2}$  &     &                          &-0.4190 \\ 
                       &$2p_{3/2}2p_{3/2}$  &     &                          &-0.1270 \\ \cline{2-5}

                       &$1d_{3/2}1d_{3/2}$  &     &                          &-1.5355 \\ 
                       &$1d_{3/2}2s_{1/2}$  &     &                          &-0.5138 \\  
                       &$1f_{7/2}1f_{7/2}$  &  2  & $^{38}$Ar$_{2.168}(2^+)$  & 0.0606 \\  
                       &$1f_{7/2}2p_{3/2}$  &     &                          & 0.0353 \\
\end{longtable}
\end{scriptsize}
\endgroup